\documentclass[prd,floatfix,onecolumn,amsmath,amssymb,10pt]{revtex4-2}
\usepackage{newtxtext,newtxmath} 
\usepackage{graphicx,color,dcolumn,booktabs,bm}
\usepackage{subfigure}
\usepackage{amssymb}
\usepackage{longtable}
\usepackage{indentfirst}
\usepackage{epsfig,gensymb,siunitx}
\usepackage{feynmf}   %{feynmp}
\usepackage{epstopdf}   %{feynmp}
\usepackage{slashed}  %for Feynman symbols
\usepackage{cases}
\usepackage{xcolor}
\definecolor{navyblue}{RGB}{0,0,170}
\definecolor{CobaltBlue}{rgb}{0,0.28,.67}
\definecolor{maroon}{RGB}{139,25,150}%burada 0-255 arasi her biri icin numara vererek renk elde et
\usepackage{multirow}
\usepackage{float}
\usepackage{pgf}
\usepackage{graphicx,color,dcolumn,booktabs,bm}
\usepackage[colorlinks, citecolor=purple,anchorcolor=purple,menucolor=purple, linkcolor=purple,filecolor=purple,runcolor=purple,urlcolor=purple,frenchlinks=purple, urlcolor=purple]{hyperref}
\usepackage{physics}
\usepackage{orcidlink}
\usepackage{bbold}
\newcommand{\mtotal}{\mathcal{P}}

\newcommand{\ga}{\gamma_\alpha}
\newcommand{\gb}{\gamma_{\beta}}
\newcommand{\s}{\sigma_{\mu\nu}}
\begin{document}
	
	\preprint{}
	
\title{\color{navyblue}{Tensor form factors of the $\Delta^+$ baryon induced by isovector and isoscalar currents in QCD}
	%Isovector and Isoscalar Tensor Form Factors of the $\Delta^+$ Baryon in QCD
}
	
	\author{Z.~Asmaee$^{1}$\orcidlink{0000-0002-3357-0574}}

	\author{N.~Hajirasouliha$^{1}$\orcidlink{0009-0004-6130-6844}}

	\author{K.~Azizi$^{1,2}$\orcidlink{0000-0003-3741-2167}}
	\email{kazem.azizi@ut.ac.ir}
	\thanks{Corresponding author}

\affiliation{
	$^{1}$Department of Physics, \href{https://ut.ac.ir/en}{University of Tehran}, North Karegar Avenue, Tehran 14395-547, Iran\\
	$^{2}$Department of Physics, \href{https://www.dogus.edu.tr/en}{Dogus University}, Dudullu-\"{U}mraniye, 34775
	Istanbul,  T\"{u}rkiye}	
\date{\today}

	\begin{abstract}
The tensor form factors of the $\Delta^+$ baryon are defined through the matrix element of the tensor current and describe its internal structure and spin distribution. We present the full Lorentz decomposition for the $\Delta^+ \rightarrow \Delta^+$ tensor current matrix element, including all independent structures consistent with Lorentz covariance, the Rarita–Schwinger constraints, and the discrete symmetries of Hermiticity, time-reversal, and parity invariance.
By investigating the tensor form factors corresponding to both the  isovector and isoscalar tensor currents, we observe differences that reflect the distinct contributions of up and down quark components in the $\Delta^+$ baryon.
	\end{abstract}
	
	\maketitle
	
	%%%%%%%%%%%%%%%%%%%%%%%%%%%%%%%%%%%%%%%%%%%%%%%%%%%%%%%%%%%%%%%%%%%%%%%%%%%%%%%%%%%
	\section{Introduction}\label{sec:one} 
One of the main challenges in non-perturbative quantum chromodynamics (QCD) is to understand the internal structure of hadrons in terms of quark and gluon degrees of freedom.
Different hadronic currents, expressed through matrix elements between hadronic states,  provide insight of their internal structure and dynamics.
The corresponding form factors serve as an effective means of describing the response of hadrons to electromagnetic, axial-vector, tensor, and gravitational currents, reflecting the distributions of charge, spin, momentum, and energy among their constituent quarks and gluons \cite{Aliev:2008cs, Er:2022cxx, Sundu:2018uyi}.
 Consequently, form factors (FFs) have received  significant attention from both experiment and theory.

In recent decades, many theoretical studies have investigated the FFs of hadrons with spins $0$ \cite{Sun:2020wfo, Fu:2023dea, Kumano:2017lhr}, $\frac{1}{2}$ \cite{Aliev:2008cs, Bincer:1959tz, Keiner:1996at, Kim:2012ts}, and $1$ \cite{Cosyn:2018thq, Sun:2017gtz, Dong:2013rk, Polyakov:2019lbq}.
The electromagnetic FFs (EMFFs) of decuplet baryons  have been studied in different approaches, such as lattice QCD \cite{Alexandrou:2008bn, Alexandrou:2010jv, Boinepalli:2009sq,lattice, Leinweber:1992hy, Lee:2005ds}, the Skyrme model \cite{Oh:1995hn}, chiral perturbation theory \cite{Geng:2009ys, Li:2016ezv}, quark models \cite{Krivoruchenko:1991pm, Berger:2004yi, Schlumpf:1993rm}, QCD sum rules (QCDSR) \cite{Aliev:2009jt, Aliev:2010uy, Aliev:2009pd, Lee:1997jk, Azizi:2009egn},  the chiral quark model \cite{Wagner:2000ii},  and chiral soliton models \cite{Kim:2019gka}.
Gravitational FFs (GFFs), associated with the energy–momentum tensor, describe the distributions of mass, momentum, and mechanical properties, such as pressure and shear forces inside hadrons. These GFFs have also been computed for decuplet baryons \cite{Kim:2020lrs, Wang:2023bjp, Dehghan:2023ytx, Dehghan:2025ncw, Dehghan:2025eov}.
Tensor form factors (TFFs) have been  investigated for octets \cite{kucukarslan:2016xhx}
and, in particular, for nucleons 
\cite{Aliev:2011ku, Gutsche:2016xff, Azizi:2019ytx, Ozdem:2020vpt, Ozdem:2021zbn}.
% However, the TFFs of decuplet baryons have not yet been studied, and this work aims to address this gap.
	
%The $\Delta$ baryon, as the lightest and lowest-lying excited state among the decuplet baryons, plays a key role in understanding the structure and dynamics of spin-$\tfrac{3}{2}$ baryons.
%Due to the short lifetime of the $\Delta$ baryon \cite{ParticleDataGroup:2022pth}, direct experimental measurements of its EMFFs remain challenging.	
%The $\Sigma^*$ and $\Xi^*$ baryons have lifetimes similar to that of the $\Delta$, while the $\Omega^-$ baryon decays more slowly \cite{ParticleDataGroup:2022pth}, which allows for more direct studies of its internal structure.
	
The tensor operator is important for studying the transversity of hadrons \cite{Ralston:1979ys, Jaffe:1991kp, Jaffe:1991ra, Barone:2001sp}.
Hadron transversity provides important information about the spin structure of quarks;  however,  its experimental determination is  challenging due to its chiral-odd nature and the lack of a direct measurement method.
The transverse parton distribution functions of the nucleons, along with the corresponding tensor charges, have also been determined using semi-inclusive deep inelastic scattering experiments \cite{Anselmino:2007fs, Anselmino:2008jk, Anselmino:2013vqa}.	
The tensor charges of hadrons have been studied theoretically using various approaches,
such as the bag model \cite{He:1994gz}, the quark models \cite{Pasquini:2005dk, Gamberg:2001qc}, and the external field method \cite{He:1996wy}.
The quark and gluon transversity generalized parton distributions (GPDs) of decuplet baryons are formulated in the light-cone gauge \cite{Fu:2024kfx}.
%Although the transversity of the nucleon has been extensively explored, that of the decuplet baryons has remained largely unexamined because of the experimental challenges associated with its measurement.	
These FFs provide essential information on the transverse spin structure of quarks  inside hadrons and allow a probabilistic description of the transverse quark densities \cite{Diehl:2005jf, Burkardt:2000za, Burkardt:2002hr}.
It has also been shown that the TFFs of hadrons can be interpreted as generalized FFs, which correspond to the Mellin moments of GPDs \cite{Fu:2024kfx, Goeke:2001tz, Diehl:2001pm, Diehl:2003ny, Belitsky:2005qn}.
Reference \cite{Fu:2024kfx} presents a decomposition of the TFFs for decuplet baryons, based on the first moments obtained from their transversity GPDs.

In this work, we present the tensor matrix element of decuplet baryons in terms of a full decomposition into TFFs, incorporating all independent structures consistent with Lorentz covariance and the Rarita–Schwinger constraints. Our approach is fully independent of the method based on transversity GPDs employed in \cite{Fu:2024kfx}, and the resulting matrix element is ensured to be Hermiticity, time-reversal (T-invariance), and parity invariant.

The $\Delta$ baryon, as the lightest and lowest-lying state among the decuplet baryons, plays a key role in understanding the structure and dynamics of spin-$\tfrac{3}{2}$ baryons.
Due to the short lifetime of the $\Delta$ baryon \cite{ParticleDataGroup:2022pth}, direct experimental measurements of its EMFFs remain challenging.	
The $\Sigma^*$ and $\Xi^*$ baryons have lifetimes similar to that of the $\Delta$, while the $\Omega^-$ baryon decays more slowly \cite{ParticleDataGroup:2022pth}, which allows for more direct studies of its internal structure.
In this article, the TFFs of the $\Delta^+$ baryon are studied using the QCDSR analysis.
The full Lorentz decomposition of the TFFs is first formulated in Sec.~\ref{QCDSR}, thereby providing the physical representation needed for the QCDSR approach in Sec.~\ref{physical side}.
Subsequently, the QCD representation of the correlation function is evaluated separately for the isovector and isoscalar cases in Sec.~\ref{QCD side}, and, the coefficients of the corresponding Lorentz structures are matched between the physical and QCD representations.
In Sec.~\ref{ANALYSES}, the behavior of the TFFs of the $\Delta^+$ baryon as a function of the squared momentum transfer $Q^2$ is analyzed numerically, and the results are extracted separately for the isovector and isoscalar cases.	
	
\section{TFFs of the $\Delta^+$ baryon in QCDSR}\label{QCDSR}
We obtain the matrix element of the tensor current between the initial and final $\Delta^+$ states in terms of ten independent TFFs, by imposing the Rarita–Schwinger constraints and enforcing Hermiticity, T-invariance, and parity (see appendix~\ref{appA}):
\begin{equation}
	\begin{split}
		\left\langle \Delta^+(p^\prime,s^\prime)\left|  \bar{\psi}(0)[i\sigma_{\mu\nu}]\psi(0)\right| \Delta^+(p,s)\right\rangle&=\bar{u}^\alpha(p^\prime,s^\prime)
		\Bigg[i\sigma_{\mu\nu} \bigg( g_{\alpha\beta}F^T_{1,0}(Q^2)+\frac{q_{\alpha}q_{\beta}}{m_\Delta^2}F^T_{1,1}(Q^2)\bigg) 
		+ig_{\alpha [\mu}\sigma _{\nu]\beta}F_{2,0}^T(Q^2)\\
		&+\frac{\gamma_{[\mu}q_{\nu]}}{m_\Delta}\bigg(g_{\alpha\beta}F_{3,0}^T(Q^2)+\frac{q_{\alpha}q_{\beta}}{m_\Delta^2}F^T_{3,1}(Q^2) \bigg) 
		+\frac{\gamma_{[\mu}\mtotal_{\nu]}\mtotal_{[\alpha}q_{\beta]}}{m_\Delta^3} F^T_{4,0}(Q^2)\\
		&+\frac{\mtotal_{[\mu}q_{\nu]}}{m_\Delta^2}\bigg(g_{\alpha\beta}F_{5,0}^T(Q^2)+\frac{q_{\alpha}q_{\beta}}{m_\Delta^2}F^T_{5,1}(Q^2) \bigg)
		+\frac{g_{\alpha[\mu}\gamma_{\nu]}q_\beta-g_{\beta[\mu}\gamma_{\nu]}q_\alpha}{m_\Delta}F^T_{6,0}(Q^2)\\
		&+\frac{g_{\alpha[\mu}q_{\nu]}\mtotal_\beta-g_{\beta[\mu}q_{\nu]}\mtotal_\alpha}{m_\Delta^2}F^T_{7,0}(Q^2)
		\Bigg]u^\beta(p,s),
		\label{martix-element}
	\end{split}
\end{equation}
where $\psi(0)$ denotes the quark Dirac field, and 
 $u^\beta(p,s)$ represents the Rarita–Schwinger spinor. $p (p')$ and $s (s')$ correspond to the momenta and spins of the initial and final states, respectively. 
The momentum transfer and total momentum are defined as 
 $q_\mu=p_\mu-p^\prime_\mu$ and $\mtotal_\mu=p_\mu+p^\prime_\mu$, with $Q^2=-q^2$. The $F^T_{i,j}(Q^2)$ represent the TFFs, $m_\Delta$ is the mass of the $\Delta^+$ baryon, and  $A_{[\mu}B_{\nu]}$ is defined as $A_\mu B_\nu-A_\nu B_\mu$.
To calculate the ten TFFs within QCDSR, we study the three-point correlation function  corresponding to the $\Delta^+\to \Delta^+$ transition induced by the tensor current,
\begin{equation}
	\Pi_{\alpha\mu\nu\beta}(p,q) = i^2 \int d^4 x e^{-ip.x} \int d^4 y e^{ip'.y}
	\langle 0 |\mathcal{T}[J_{\alpha}^{\Delta}(y)J^T_{\mu \nu}(0)\bar{J}_{\beta}^{\Delta}(x)]| 0 \rangle,
	\label{corrf}
\end{equation}
where $\mathcal{T}$ represents the time-ordering operator, and  $J^T_{\mu\nu}$ denotes the tensor current. The interpolating current $J^{\Delta}_\alpha(y)$, which represents the $\Delta^+(uud)$ state at the space–time point $y$ with Lorentz index $\alpha$, is defined explicitly as \cite{Dehghan:2023ytx},
\begin{equation}
	\begin{split}
&J_{\alpha}^{\Delta}(y)=\frac{1}{\sqrt{3}}\varepsilon^{abc}\left[
	\vphantom{\int_0^{y_2}}
	2 \Big(u^{aT} (y) C\gamma_{\alpha}d^{b} (y) \Big)  u^{c} (y) +
	\Big(u^{aT} (y) C\gamma_{\alpha}u^{b} (y) \Big) d^{c} (y) \right],
	\end{split}
	\label{J-ab}
\end{equation}
where $C$ denotes the charge-conjugation operator, and $a, b$ and $c$ label the color indices.
 The tensor current $J^T_{\mu\nu}(0)=\bar{\psi}(0)i\sigma_{\mu\nu} \psi(0)$ for $\Delta^+$ is chosen as \cite{Aliev:2011ku},
\begin{equation}
	J^T_{\mu\nu}(0)=\bar{u}(0)i\sigma_{\mu\nu} u(0)\pm \bar{d}(0)i\sigma_{\mu\nu} d(0),
	\label{tensor-current}
\end{equation}
where the upper and lower signs correspond to the isosinglet and isovector cases, respectively.

The QCDSR for the FFs are obtained by evaluating  Eq.~\eqref{corrf} in two equivalent representations: the physical side, expressed in terms of hadronic parameters, and the QCD side, formulated in terms of quark and gluon degrees of freedom.
 Double Borel transformations with respect to $p^2$ and $p^{\prime 2}$, along with continuum subtraction, are applied to both sides to suppress the excited state contributions and enhance those of the ground state pole.
 Finally, the TFFs are extracted by equating coefficients of identical Lorentz structures in the physical and QCD representations.

\subsection{Physical side}\label{physical side}
Since the interpolating current acts as the operator that creates or annihilates the $\Delta^+$ states under study from the vacuum, two complete sets of intermediate states carrying the same quantum numbers as the $\Delta^+$ baryon are inserted to evaluate Eq.~\eqref{corrf} in terms of hadronic parameters \cite{Dehghan:2023ytx},
\begin{equation}
	\begin{split}
		&1 = \left| 0\left\rangle \right\langle 0\right| +\sum_l \int \frac{d^4p_l}{(2\pi)^4} (2\pi) \delta(p_l^2 - m^2_\Delta) \left| \Delta(p_l)\rangle \langle \Delta(p_l)\right| + \text{higher Fock states},
		\label{set complet}
	\end{split}
\end{equation}
where $\Delta(p_l)$  denotes the $\Delta^+$ state with four-momentum $p_l$.
After performing the necessary calculations and the four-dimensional integrations over the space-time coordinate, and separating the ground-state contribution from those of the excited states, the three-point correlation function is finally obtained as:
\begin{equation}
	\Pi_{\alpha\mu\nu\beta}^\text{Had}(p,p^\prime) = 
	\sum_{s,s{'}}\frac{\langle0|J_{\alpha}^{\Delta}(0)|{\Delta(p',s')}\rangle\langle {\Delta(p',s')}
		|J^T_{\mu \nu}(0)|
		{\Delta(p,s)}\rangle\langle {\Delta(p,s)}
		|\bar{J}_{\beta}^{\Delta}(0)| 0 \rangle}
	{(m^2_{\Delta}-p'^2)(m^2_{\Delta}-p^2)} 
	+\cdots,
	\label{physicalside correlation function}
\end{equation}
where the dots indicate contributions from higher states and the continuum.
The matrix elements of the interpolating current $J_{\alpha}^{\Delta}(0)$ between the vacuum and $\Delta^+$ baryon are written as \cite{Aliev:2011ku, Ioffe:1981kw, Aliev:2002ra, Azizi:2014yea},
\begin{equation}
	\begin{split}
		&\langle0|J_{\alpha}^{\Delta}(0)|{\Delta^+(p',s')}\rangle=\lambda_\Delta u_{\alpha}(p^\prime, s^\prime),
		\label{residue}
	\end{split}
\end{equation}	
where $\lambda_\Delta$ is the residue of the $\Delta^+$ baryon.	
The sum over Rarita–Schwinger spinors for the $\Delta^+$ baryon is given by \cite{Dehghan:2023ytx},
	\begin{equation}
		\begin{split}
	&	\sum_{s'}{u_\alpha}(p',s') \bar{u}_{\alpha'}(p',s') = 
		- ({\slashed{p}^\prime} + m_\Delta)
		\Big[g_{\alpha\alpha'} - \frac{\gamma_{\alpha}\gamma_{\alpha'}}{3}
		-\frac{2 p'_{\alpha} p'_{\alpha'}}{3 m_\Delta^2}
		+\frac{p'_{\alpha} \gamma_{\alpha'} - p'_{\alpha'} \gamma_{\alpha}}{3 m_\Delta}
		\Big].
		\end{split}
		\label{sum spin}
	\end{equation}
	By substituting Eqs.~\eqref{martix-element}, \eqref{residue}, and \eqref{sum spin} into Eq.~\eqref{physicalside correlation function}, we obtain the physical representation of the correlation function for the $\Delta^+ \to \Delta^+$  transition.
It is important to note that  the interpolating current $J^\Delta_\alpha$ is intended to couple to $\Delta^+$ baryon with spin-$\frac{3}{2}$, but due to its Lorentz structure, it can also have a nonzero overlap with spin-$\frac{1}{2}$ states. This overlap leads to unwanted contributions, which can be written as \cite{Dehghan:2023ytx},
	\begin{equation}
		\begin{split}
&	\langle0\mid J^\Delta_{\alpha}(0)\mid p', s'=1/2\rangle=(A  p^\prime_{\alpha}+B\gamma_{\alpha})u(p^\prime,s^\prime=1/2),
	\end{split}
		\label{unwnted}
	\end{equation}
	where $A$ and $B$ are scalar functions and $u(p', s'=1/2)$ is  the Dirac spinor.
From Eq.~\eqref{unwnted}, the spin-$\frac{1}{2}$ contamination, which appears through terms proportional to $\gamma_{\alpha}$ and $p'_\alpha$, can be eliminated by imposing the constraints
\begin{equation}
	\begin{split}
		&\gamma^\alpha J_\alpha^\Delta=0,\quad p^{\prime \alpha} J_\alpha^\Delta=0,
	\end{split}
\end{equation}
allowing the coefficient $A$ to be expressed in terms of $B$ and effectively projecting out the unwanted spin-$\frac{1}{2}$ components. As a result, the pure spin-$\frac{3}{2}$ contributions in the correlation function are isolated.
In practice, this is implemented by first arranging the Dirac matrices in the order
 $\gamma_{\alpha}\slashed{p}^\prime\slashed{p}\gamma_{\mu} \gamma_{\nu}\gamma_{\beta}$, and then removing terms that begin with $\gamma_{\alpha}$, end with $\gamma_{\beta}$, or are proportional to $p^\prime_\alpha$ and $p_\beta$.
Finally, the double Borel transformation with respect to $p^2$ and $p^{\prime2}$ is applied using the Borel parameters $M_1^2$ and $M^2_2$ \cite{Ozdem:2017jqh, Azizi:2018duk}, respectively, 
\begin{equation}
		\mathcal{B}_{M_1^2}	\mathcal{B}_{M_2^2}\left( \frac{1}{\left( m_\Delta^2-p^{2}\right) \left( m_\Delta^2-p^{\prime 2}\right) }\right) =e^{-\frac{m_\Delta^2}{M_1^2}}e^{-\frac{m_\Delta ^2}{M_2^2}}=e^{-\frac{m_\Delta^2}{M^2}}.
		\label{Borel}
\end{equation}
In the second equality of Eq.~\eqref{Borel}, we use the fact that both the initial and final states correspond to the same $\Delta^+$ baryons; therefore, the Borel masses were taken to be equal,  $M_1^2=M_2^2=2M^2$.
After applying these steps, the hadronic side is obtained in its final form, in Borel scheme,  as:
			\begin{equation}\begin{split}
			\Pi_{\alpha\mu\nu\beta}^\text{Had}(M^2,Q^2) = \left|\lambda_\Delta \right|^2 e^{-\frac{m_\Delta^2}{M^2}}&\bigg[ g_{\alpha\beta}\gamma_{\mu}\gamma_\nu\slashed{p}^\prime \slashed{p}\Pi_1^\text{Had}(Q^2)
			+g_{\alpha\mu}g_{\nu\beta}\slashed{p}^\prime \slashed{p}\Pi_2^\text{Had}(Q^2)
			+ p_\alpha p^\prime_\beta \gamma_{\mu}\gamma_\nu\slashed{p}^\prime \slashed{p}\Pi_3^\text{Had}(Q^2)\\
			&
			+ g_{\alpha\beta}\gamma_\mu p^\prime_\nu\slashed{p}^\prime \slashed{p}\Pi_4^\text{Had}(Q^2)+p_\mu g_{\alpha\beta}p^\prime_\nu \slashed{p}^\prime\slashed{p}\Pi_5^\text{Had}(Q^2)+p_\alpha g_{\mu\beta}\gamma_\nu \slashed{p}^\prime \slashed{p}\Pi_6^\text{Had}(Q^2)\\&
			+p_\alpha p^\prime_\mu g_{\nu\beta} \slashed{p}^\prime\Pi_7^\text{Had}(Q^2)+p_\alpha p_\mu p^\prime_\nu p^\prime_\beta \slashed{p}^\prime\Pi_{8}^\text{Had}(Q^2)
			+p_{\alpha}p^\prime_\mu p_\nu {p}^\prime_\beta\Pi_{9}^\text{Had}(Q^2)\\
			&
			+p_{\alpha} \gamma_\mu p_\nu p^\prime_\beta \slashed{p}^\prime \Pi_{10}^\text{Had}(Q^2)+p_\alpha p_\mu \gamma_\nu p^\prime_\beta \slashed{p} \Pi_{11}^\text{Had}(Q^2)
			+p_\alpha p^\prime_\mu \gamma_\nu p^\prime_\beta \slashed{p}\Pi_{12}^\text{Had}(Q^2)
			\\
			&+g_{\alpha\beta}p_\mu p^\prime_\nu \slashed{p}^\prime\Pi_{13}^\text{Had}(Q^2)
			+\cdots\bigg],
		\end{split}
		\label{physical side1}
		\end{equation}
where the functions $\Pi_ {i}^\text{Had}(Q^2)$ ($ i $ runs from $ 1 $	to $ 13 $) depend on the TFFs as well as other hadronic parameters (see appendix~\ref{appB}).
Only the Lorentz structures relevant for the isoscalar and isovector TFFs are kept.  
Here, the dots denote the contributions from higher states and the continuum, as well as other structures involved.

We should point out that although the tensor current matrix element Eq.~\eqref{martix-element} is antisymmetric under $\mu\leftrightarrow\nu$, the ordering of Dirac matrices required to eliminate spin-$\frac{1}{2}$ contamination can  generate symmetric terms such as $g_{\mu\nu}$. While this may appear to violate antisymmetry, it does not pose any problem, since these terms do not enter the matching with the Lorentz structures on the QCD side, and their contributions are already accounted for in the dots in Eq.~\eqref{physical side1}.
	\subsection{QCD side}\label{QCD side}
In this section, we express the correlation function in terms of quark and gluon degrees of freedom. Substituting the tensor current from Eq.~\eqref{tensor-current} and the interpolating currents from Eq.~\eqref{J-ab} into Eq.~\eqref{corrf}, and taking into account the time-ordering operator in its definition, we apply Wick’s theorem. As a result, the correlation function on the QCD side is obtained as
\begin{equation}
	\Pi_{\alpha\mu\nu\beta}^\text{QCD}(p,p^\prime) = i^2\varepsilon^{abc}\varepsilon^{a'b'c'}
	\int d^4 x\ e^{-ip.x} \int d^4 y\ e^{ip'.y}\ \Pi_{\alpha\mu\nu\beta}(x,y),
	\label{corrf1}
\end{equation}
where $\Pi_{\alpha\mu\nu\beta}(x,y)$ is formulated in terms of the light-quarks propagators and the Dirac gamma matrices.
 Since both isovector and isoscalar tensor currents are considered, the QCD side is represented separately for each current type.
As they are rather lengthy, the full expressions for both cases are provided in Appendix~\ref{appC}.
Substituting Eqs.~\eqref{isoscalar}, \eqref{isovector} and \eqref{propagator} into Eq.~\eqref{corrf1}, the correlation function is obtained as	
\begin{equation}
\Pi_{\alpha\mu\nu\beta}^\text{QCD}(p,p')=\Pi^{(\text{pert})}_{\alpha\mu\nu\beta}(p,p')+\sum_{d=3}^{6}\Pi^{d}_{\alpha\mu\nu\beta}(p,p')+\cdots,
\label{corrf2}
\end{equation}	
where the first term corresponds to the perturbative contribution (with operator mass dimension $ d=0 $), while the second term represents the nonperturbative part arising from operators of mass dimension three to six. Contributions from operators of higher mass dimensions are also included in the dots.
The terms with mass dimensions $0$, $3$ and $4$ and those with mass dimensions $5$ and $6$ were calculated using two different methods.

\underline{\textbf{(i) Terms with operator  mass dimensions $0$, $3$ and $4$:}} Eq.~\eqref{corrf2} is derived in coordinate space and subsequently evaluated in momentum space for these terms using \cite{Azizi:2017ubq}
	\begin{equation}
		\dfrac{1}{(L^2)^{m_j}} = \int \frac{d^D p_j}{(2 \pi)^D} \exp{-ip_j.L} 
		\,i (-1)^{m_j + 1} 2^{D - 2 m_j} \pi^{D/2} 
		\dfrac{\Gamma[D/2 - m_j]}{\Gamma[m_j]} {\Big(-\frac{1}{p_j^2}\Big)}^{D/2 - m_j},
		\label{fourie}
	\end{equation}
where $L$ denotes $x-y$, $x$ or $y$. The coordinates are represented as derivatives with respect to momenta $p'$ and $p$: $y_\mu=-i \frac{\partial}{ {\partial p'_{\mu}}}$ and  $x_\mu=i \frac{\partial}{{\partial p_{\mu}}}$.
Integrating over $y$ and $x$ in $D$-dimensions yields two momentum Dirac delta, which allow two of the $D$-dimensional momentum integrals in Eq.~\eqref{fourie} to be evaluated directly.  As a result, a $D$-dimensional momentum integral remains, which is first simplified using the Feynman parametrization and then evaluated using the following formula \cite{Dehghan:2023ytx, Azizi:2017ubq},
\begin{equation}
	\int d^D\ell \frac{1}{(\ell^2 + \Delta)^n} = 
	\dfrac{i \pi^{D/2} (-1)^{n}\Gamma[n-D/2]}{\Gamma[n] (-\Delta)^{n-D/2}}.
	\label{integral}
\end{equation}
According to the dispersion relation,  the invariant amplitudes corresponding to the  coefficients of different Lorentz structures in QCD side, can be expressed in terms of the spectral densities as a double dispersion integral in momentum space,
\begin{equation}
\Pi_{i}^{\text{QCD}}(s_0, Q^2) = \int_{(2m_u+m_d)^2}^{s_0} ds \int_{(2m_u+m_d)^2}^{s_0} ds' \frac{\rho_i(s,s',Q^2)}{(s-p^2)(s'-p'^2)},
\label{spectral density}
\end{equation}
where $s_0$ represents the continuum threshold. 
The spectral densities $\rho_i(s,s',Q^2)$ are defined by the imaginary parts of $\Pi_{i}^{\text{QCD}}(Q^2)$	as $	\rho_i(s,s',Q^2) =\text{Im}[\Pi_{i}^{\text{QCD}}(s_0,Q^2)]/\pi$ and 
the imaginary parts corresponding to different structures  can be calculated using \cite{Azizi:2017ubq},
\begin{equation}
	\Gamma[\frac{D}{2} - n] {\Big(\frac{-1}{\Delta}\Big)}^{D/2 - n} = 
	\frac{(-1)^{n-1}}{(n-2)!} (-\Delta)^{n-2} \ln [-\Delta].
	\label{imarinarypart}
\end{equation}
In the next step, we set $D=4$, corresponding to four dimensional space-time.  As on the physical side, the Dirac matrices are arranged in the same order, after which the necessary constraints to remove the spin-$1/2$ contamination are applied.
	
\underline{\textbf{(ii) Terms with operator mass dimensions $5$ and  $6$:}}
For these terms, we first perform a Wick rotation from Minkowski to Euclidean space, followed by the application of Schwinger parametrization as
\begin{equation}
	\dfrac{1}{(L^2)^{m_j}}= \frac{(-1)^{m_j}}{\Gamma(m_j)} \int_0^\infty d\alpha\, \alpha^{m_j-1} \exp{-\alpha L^2},
	\label{Schwinger}
\end{equation}
where $\alpha$ is an independent Schwinger parameter. We perform the Gaussian integration over $x_E$ and $y_E$, whose result is expressed in terms of the Schwinger parameters. Next, we set $D=4$ and apply a Wick rotation back to Minkowski space to implement $x_\mu$ and $y_\mu$ as derivatives with respect to the momenta. As on the physical side, the Dirac matrices are then ordered in the same method, and the spin-$1/2$ contamination is removed. In the following step, we perform a double Borel transformation with respect to the squares of the initial and final momenta, using the Borel parameters $M_1^2$ and $M_2^2$ (see Eq.~\ref{borel}), and then the integration over the Schwinger parameters is carried out.

Finally, after performing a double Borel transformation for part $\textbf{(i)}$ and applying continuum subtraction for both parts $\textbf{(i)}$ and $\textbf{(ii)}$ \cite{Ozdem:2017jqh, Azizi:2018duk}, the QCD side of the correlation function takes the following form in Borel scheme:
\begin{equation}
	\Pi_i^{\text{QCD}}(s_0, M^2, Q^2) = \int_{(2m_u+m_d)^2}^{s_0} ds \int_{(2m_u+m_d)^2}^{s_0} ds' \rho_i(s,s',Q^2)\ e^{-\frac{s}{2M^2}}\ e^{-\frac{s^\prime}{2M^2}}+\Gamma_i(s_0,M^2, Q^2),
	\label{spectral density1}
\end{equation}
where, in the above equation,  the first term corresponds to the result of part $\textbf{(i)}$, while the second term, $\Gamma_i(s_0,M^2, Q^2)  $, corresponds to the result of part $\textbf{(ii)}$.
In its explicit form, the QCD side is obtained as:
		\begin{equation}\begin{split}
			\Pi_{\alpha\mu\nu\beta}^\text{QCD}(s_0, M^2, Q^2) &= g_{\alpha\beta}\gamma_{\mu}\gamma_\nu\slashed{p}^\prime \slashed{p}\Pi_1^\text{QCD}(s_0, M^2, Q^2)
			+g_{\alpha\mu}g_{\nu\beta}\slashed{p}^\prime \slashed{p}\Pi_2^\text{QCD}(s_0, M^2, Q^2)
			+ p_\alpha p^\prime_\beta \gamma_{\mu}\gamma_\nu\slashed{p}^\prime \slashed{p}\Pi_3^\text{QCD}(s_0, M^2, Q^2)\\
			&
			+ g_{\alpha\beta}\gamma_\mu p^\prime_\nu\slashed{p}^\prime \slashed{p}\Pi_4^\text{QCD}(s_0, M^2, Q^2)+p_\mu g_{\alpha\beta}p^\prime_\nu \slashed{p}^\prime\slashed{p}\Pi_5^\text{QCD}(s_0, M^2, Q^2)+p_\alpha g_{\mu\beta}\gamma_\nu \slashed{p}^\prime \slashed{p}\Pi_6^\text{QCD}(s_0, M^2, Q^2)\\&
			+p_\alpha p^\prime_\mu g_{\nu\beta} \slashed{p}^\prime\Pi_7^\text{QCD}(s_0, M^2, Q^2)+p_\alpha p_\mu p^\prime_\nu p^\prime_\beta \slashed{p}^\prime\Pi_{8}^\text{QCD}(s_0, M^2, Q^2)
			+p_{\alpha}p^\prime_\mu p_\nu {p}^\prime_\beta\Pi_{9}^\text{QCD}(s_0, M^2, Q^2)\\
			&
			+p_{\alpha} \gamma_\mu p_\nu p^\prime_\beta \slashed{p}^\prime \Pi_{10}^\text{QCD}(s_0, M^2, Q^2)+p_\alpha p_\mu \gamma_\nu p^\prime_\beta \slashed{p} \Pi_{11}^\text{QCD}(s_0, M^2, Q^2)
			+p_\alpha p^\prime_\mu \gamma_\nu p^\prime_\beta \slashed{p}\Pi_{12}^\text{QCD}(s_0, M^2, Q^2)
			\\
			&+g_{\alpha\beta}p_\mu p^\prime_\nu \slashed{p}^\prime\Pi_{13}^\text{QCD}(s_0, M^2, Q^2)
			+\cdots.
		\end{split}
		\label{QCD side1}
	\end{equation}
	The functions $\Pi^{\text{QCD}}_i(s_0, M^2, Q^2)$ involve very lengthy expressions, and therefore their explicit forms are not presented.
	The TFFs of the $\Delta^+$ baryon are extracted by matching the coefficients of the selected Lorentz structures between the QCD side (Eq.~\eqref{QCD side1}) and the physical side (Eq.~\eqref{physical side1}).
	
\section{ NUMERICAL ANALYSES}\label{ANALYSES}	
	The TFFs derived from the QCDSR, as discussed in the preceding sections, are analyzed numerically in this section. The values of the relevant input parameters are:
	$m_u=0.00216\pm0.00007\ \text{GeV}$, 	$m_d=0.00470\pm0.00007\ \text{GeV}$ and $m_\Delta=1.2349\pm0.0014\ \text{GeV}$ \cite{ParticleDataGroup:2024cfk}. The two-quark, two-gluon, and mixed quark–gluon condensates are taken as  $\left\langle \bar{u}u \right\rangle=\left\langle \bar{d}d \right\rangle =(-0.24\pm 0.01)^3$ $\mathrm{GeV}^3$ \cite{Belyaev:1982sa}, $\langle \frac{\alpha_s}{\pi} G^2 \rangle   = (0.012\pm0.004)$ $~\mathrm{GeV}^4 $ \cite{Belyaev:1982cd} and 
	$\left\langle \bar{q} g_s \sigma G q\right\rangle  = (0.8 \pm 0.1)\left\langle \bar{q}q \right\rangle \ \mathrm{GeV}^5$ ($q=u,d$) \cite{Belyaev:1982sa}, respectively.
	The  residue, the strong coupling constant and the strong gauge coupling are $\lambda_\Delta= 0.038~\mathrm{GeV}^3$ \cite{Aliev:2007pi}, $\alpha_s=(0.118 \pm 0.005)$ \cite{DELPHI:1993ukk} and $g_s=\sqrt{4\pi \alpha_s}$, respectively.

\begin{figure}[!htb]
		\centering
		\includegraphics[width=0.38\textwidth]{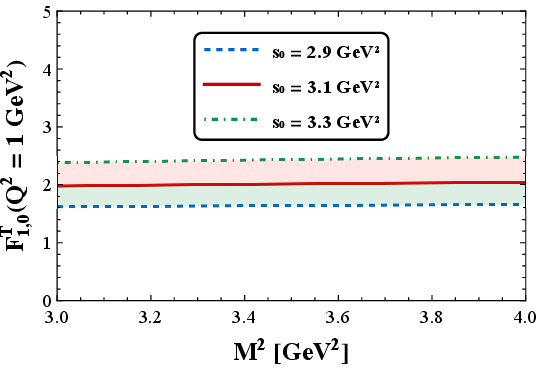}~~~~~~~~
		\includegraphics[width=0.38\textwidth]{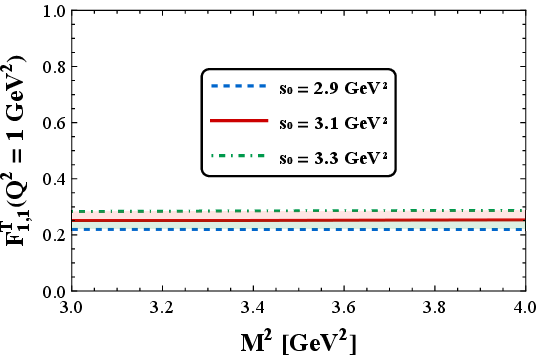}
		
		\vspace{0.1cm}
		\includegraphics[width=0.38\textwidth]{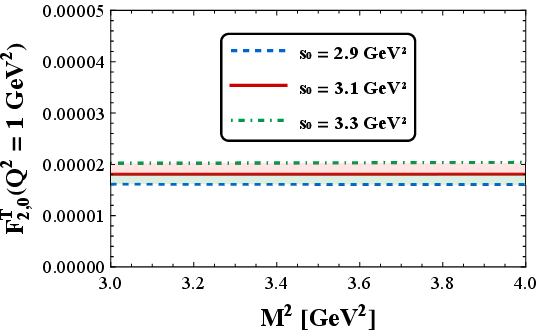}~~~~~~~~
		\includegraphics[width=0.38\textwidth]{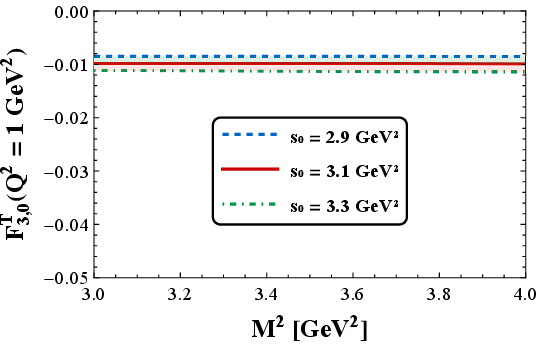}
		
		\vspace{0.1cm}
		\includegraphics[width=0.38\textwidth]{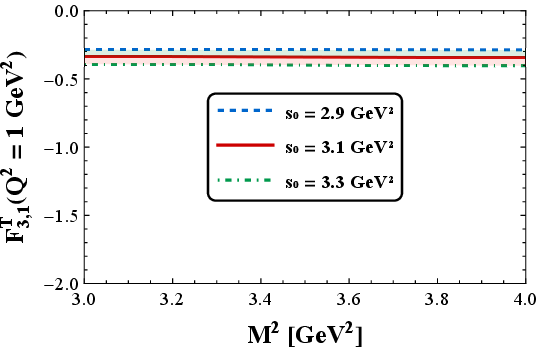}~~~~~~~~
		\includegraphics[width=0.38\textwidth]{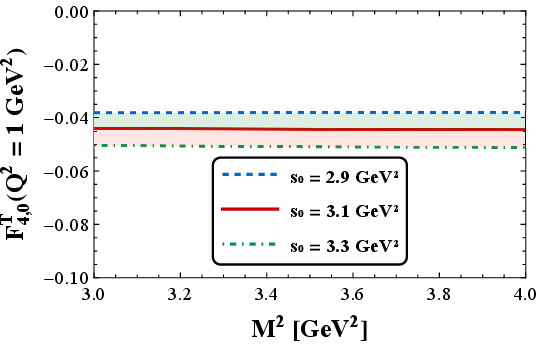}
		
		\vspace{0.1cm}
		\includegraphics[width=0.38\textwidth]{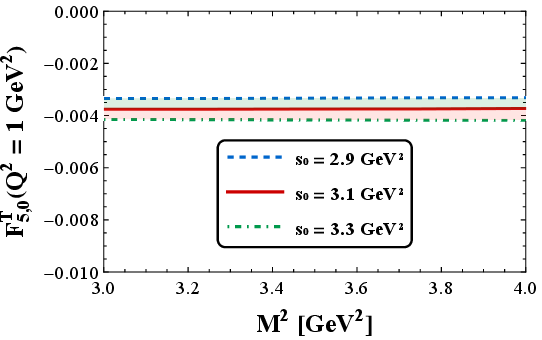}~~~~~~~~
		\includegraphics[width=0.38\textwidth]{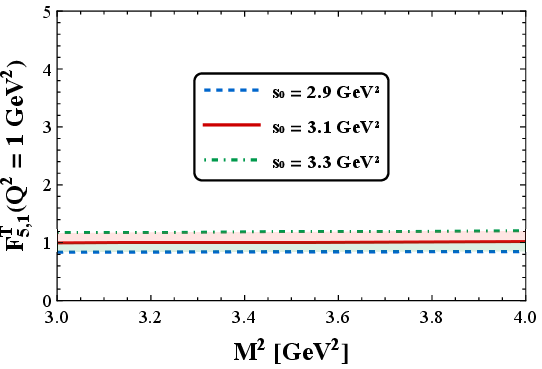}
		
		\vspace{0.1cm}
		\includegraphics[width=0.38\textwidth]{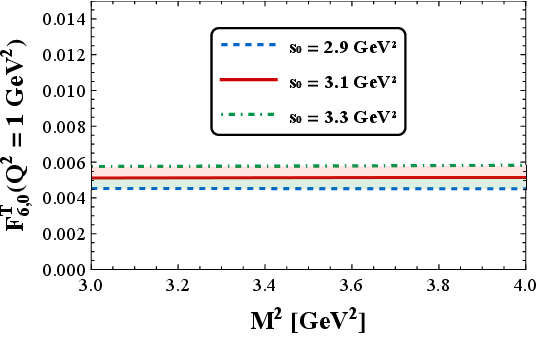}~~~~~~~~
		\includegraphics[width=0.38\textwidth]{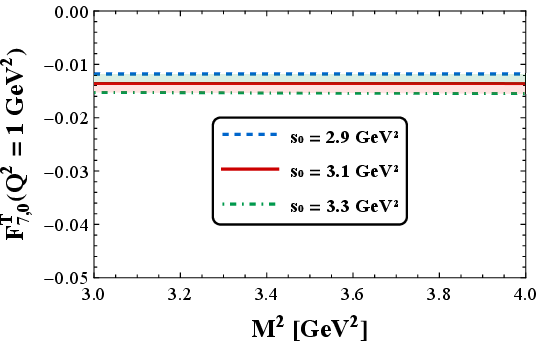}
		\caption{The $M^2$ dependence of the isoscalar TFFs of the $\Delta$ baryon at $Q^2 = 1.0~\text{GeV}^2$ for three values of the continuum threshold $s_0$.}
		\label{Msqisoscalar}
	\end{figure}

\begin{figure}[!htb]
		\centering
		\includegraphics[width=0.4\textwidth]{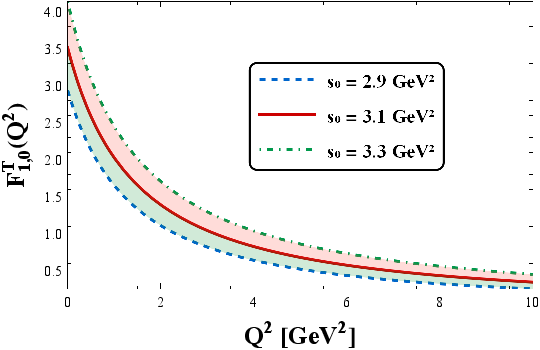}~~~~~~~~
		\includegraphics[width=0.4\textwidth]{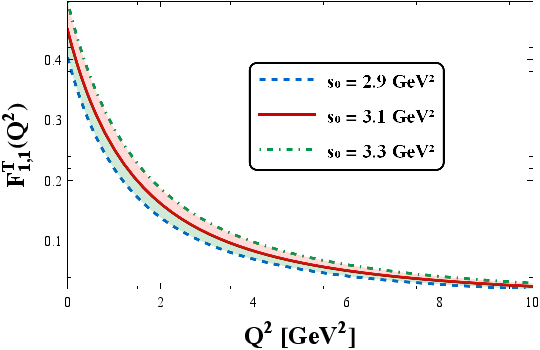}
		
		\vspace{0.1cm}
		\includegraphics[width=0.4\textwidth]{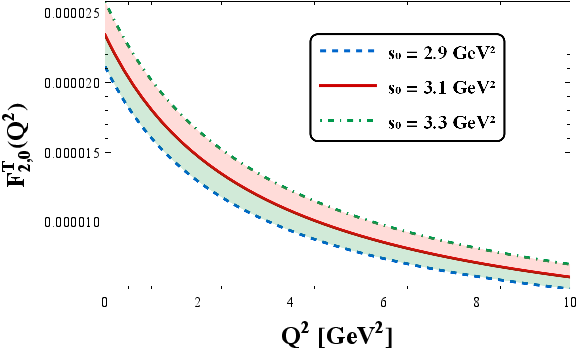}~~~~~~~~
		\includegraphics[width=0.4\textwidth]{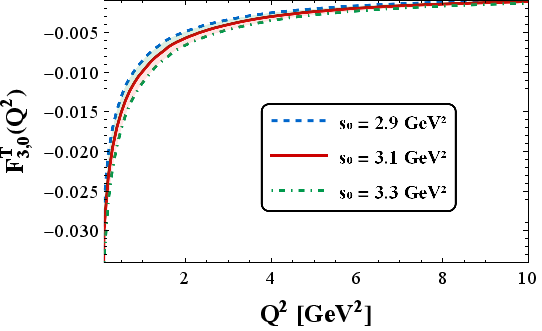}
		
		\vspace{0.1cm}
		\includegraphics[width=0.4\textwidth]{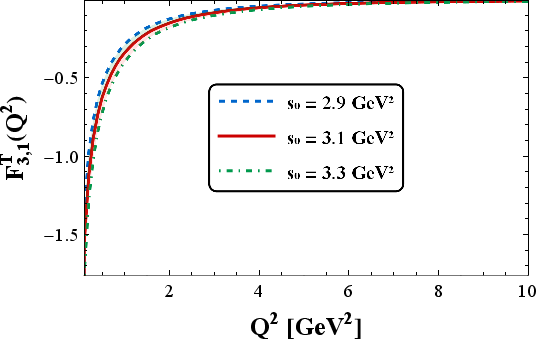}~~~~~~~~
		\includegraphics[width=0.4\textwidth]{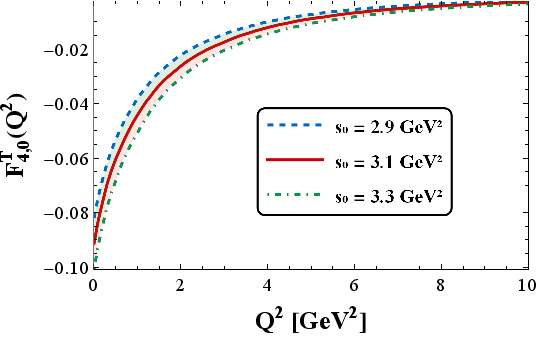}
		
		\vspace{0.1cm}
		\includegraphics[width=0.4\textwidth]{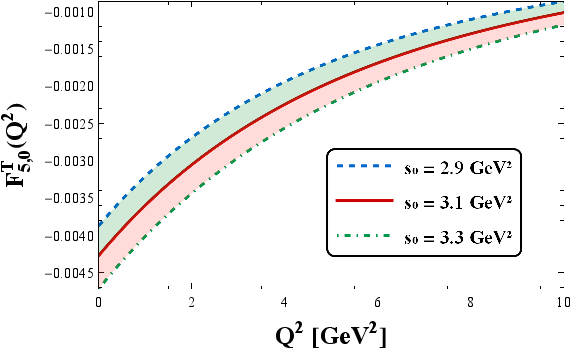}~~~~~~~~
		\includegraphics[width=0.4\textwidth]{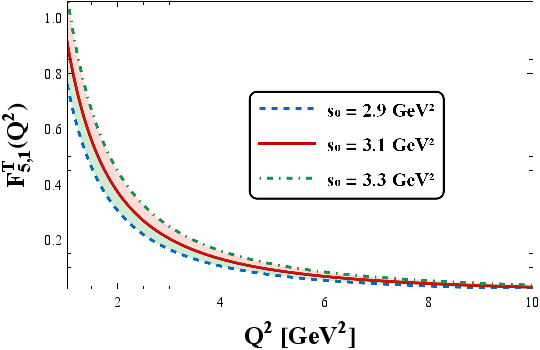}
		
		\vspace{0.1cm}
		\includegraphics[width=0.4\textwidth]{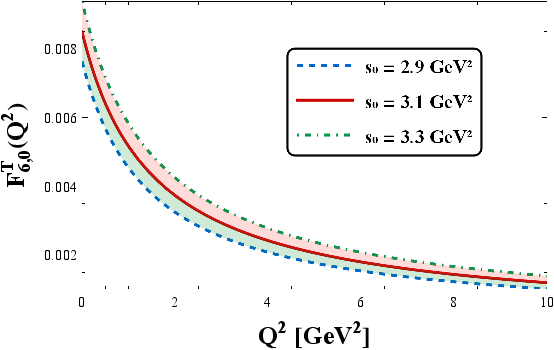}~~~~~~~~
		\includegraphics[width=0.4\textwidth]{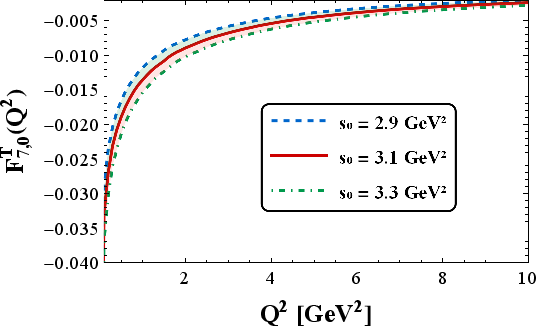}
		\caption{The $Q^2$ dependence of the isoscalar TFFs of the $\Delta$ baryon at $M^2 = 3.5~\text{GeV}^2$ for three values of the continuum threshold $s_0$.}
		\label{Qsqisoscalar}
	\end{figure}

\begin{figure}[!htb]
		\centering
		\includegraphics[width=0.38\textwidth]{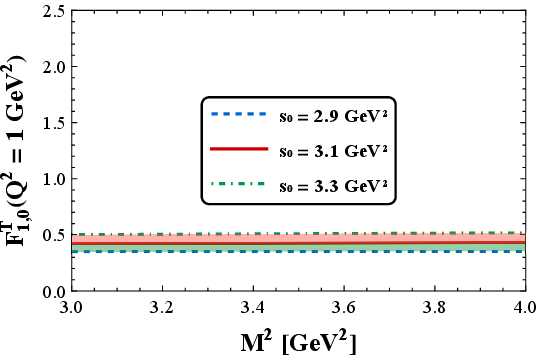}~~~~~~~~
		\includegraphics[width=0.38\textwidth]{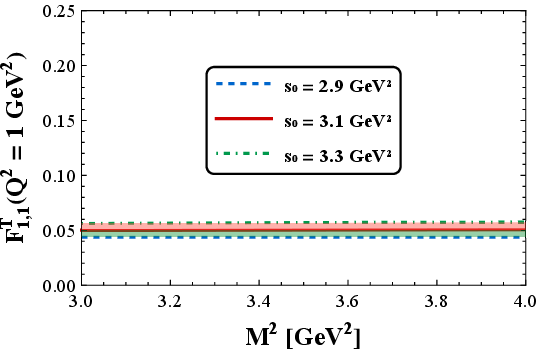}
		
		\vspace{0.1cm}
		\includegraphics[width=0.38\textwidth]{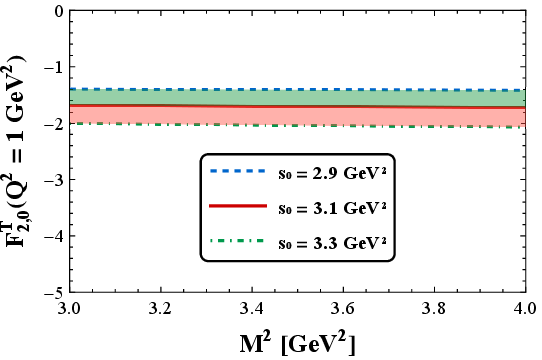}~~~~~~~~
		\includegraphics[width=0.38\textwidth]{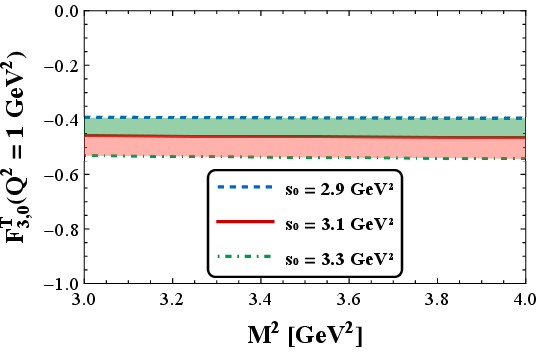}
		
		\vspace{0.1cm}
		\includegraphics[width=0.38\textwidth]{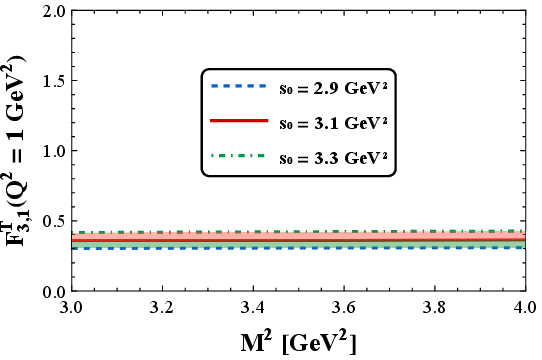}~~~~~~~~
		\includegraphics[width=0.38\textwidth]{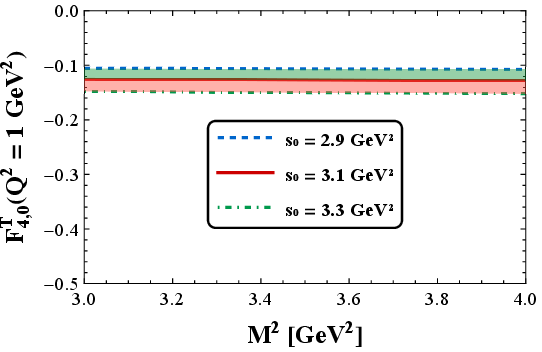}
		
		\vspace{0.1cm}
		\includegraphics[width=0.38\textwidth]{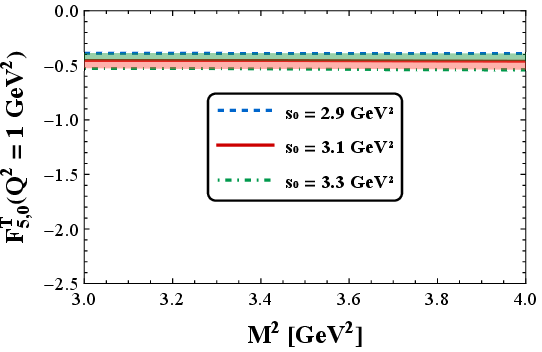}~~~~~~~~
		\includegraphics[width=0.38\textwidth]{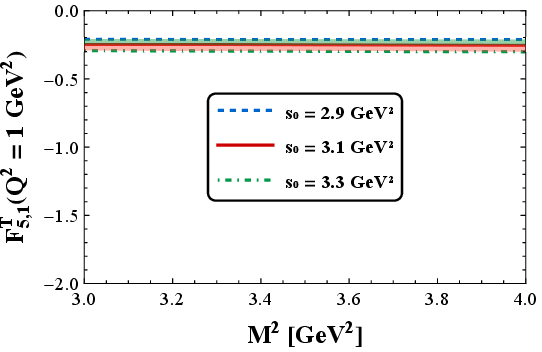}
		
		\vspace{0.1cm}
		\includegraphics[width=0.38\textwidth]{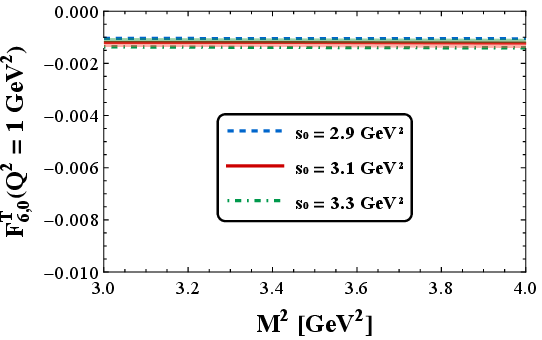}~~~~~~~~
		\includegraphics[width=0.38\textwidth]{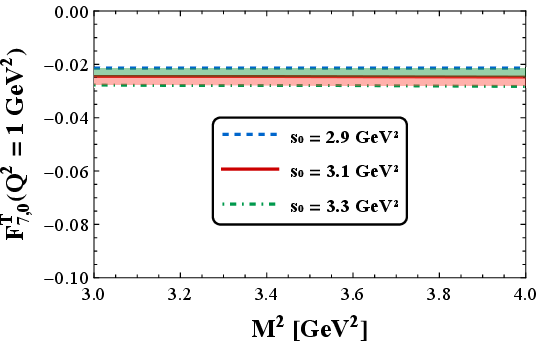}
		\caption{The $M^2$ dependence of the isovector TFFs of the $\Delta$ baryon at $Q^2 = 1.0~\text{GeV}^2$ for three values of the continuum threshold $s_0$.}
		\label{Msqisovector}
	\end{figure}

\begin{figure}[!htb]
		\centering
		\includegraphics[width=0.4\textwidth]{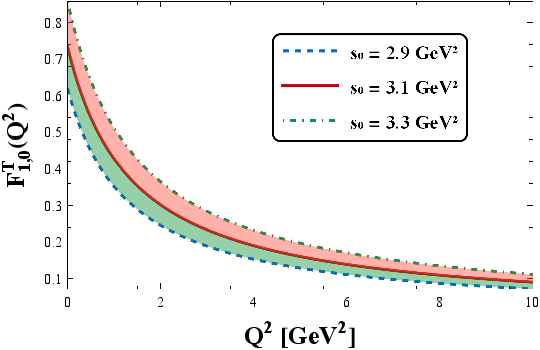}~~~~~~~~
		\includegraphics[width=0.4\textwidth]{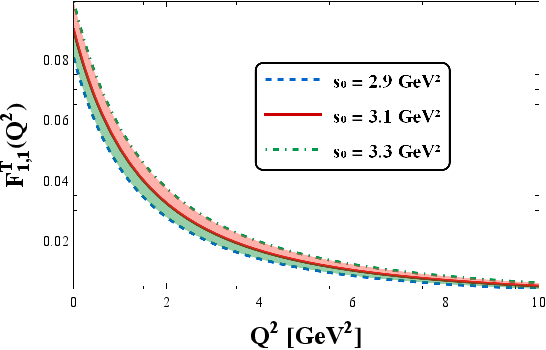}
		
		\vspace{0.1cm}
		\includegraphics[width=0.4\textwidth]{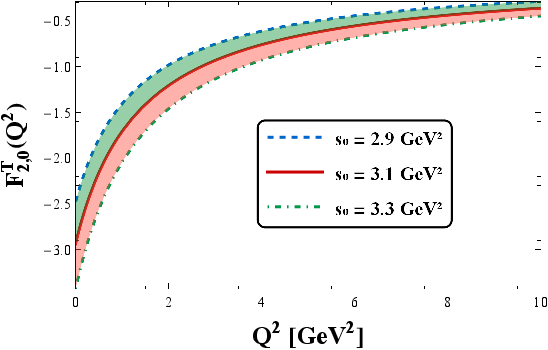}~~~~~~~~
		\includegraphics[width=0.4\textwidth]{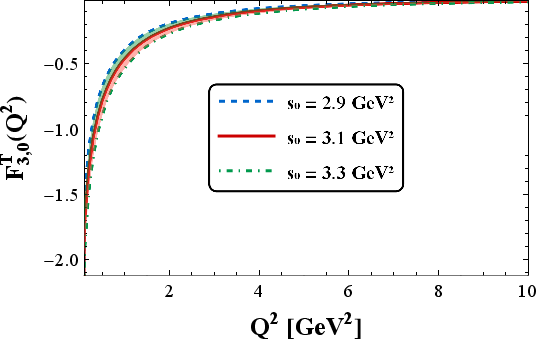}
		
		\vspace{0.1cm}
		\includegraphics[width=0.4\textwidth]{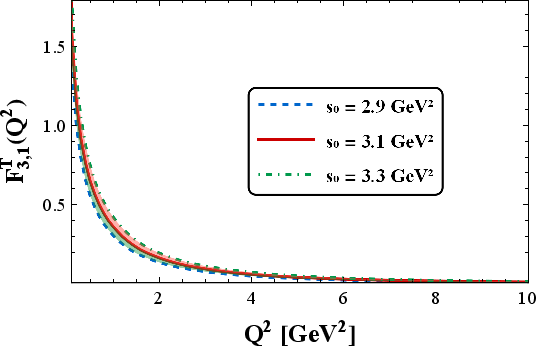}~~~~~~~~
		\includegraphics[width=0.4\textwidth]{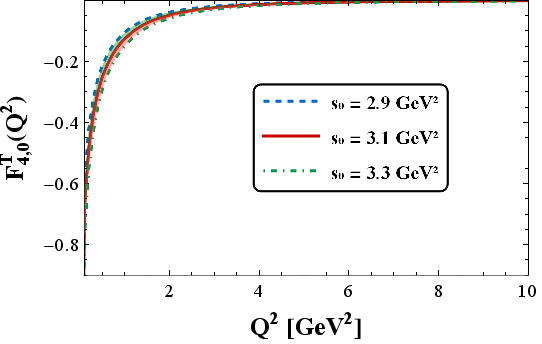}
		
		\vspace{0.1cm}
		\includegraphics[width=0.4\textwidth]{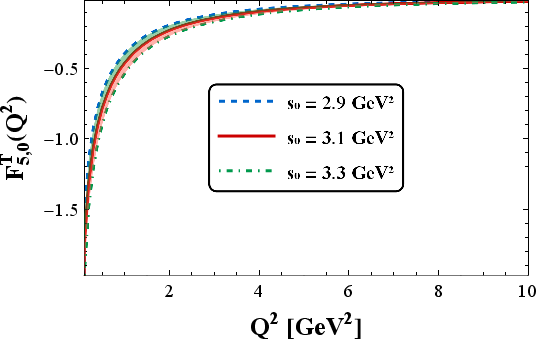}~~~~~~~~
		\includegraphics[width=0.4\textwidth]{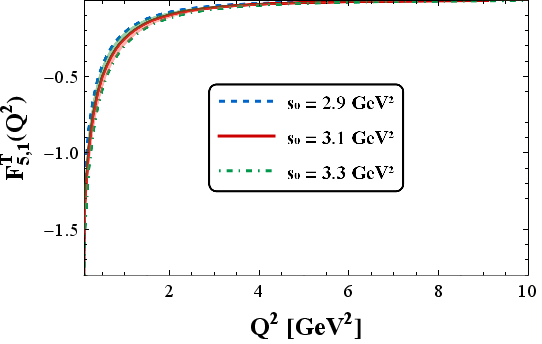}
		
		\vspace{0.1cm}
		\includegraphics[width=0.4\textwidth]{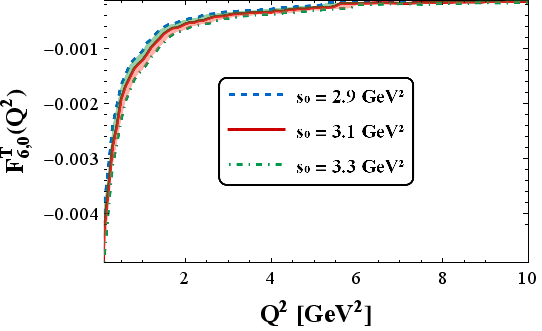}~~~~~~~~
		\includegraphics[width=0.4\textwidth]{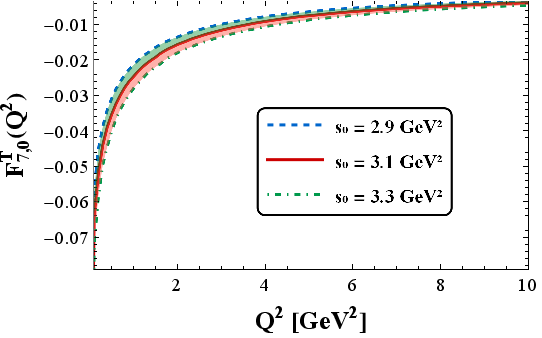}
		\caption{The $Q^2$ dependence of the isovector TFFs of the $\Delta$ baryon at $M^2 = 3.5~\text{GeV}^2$ for three values of the continuum threshold $s_0$.}
		\label{Qsqisovector}
	\end{figure}

Besides these input parameters, the QCDSR also involve two auxiliary quantities: the Borel mass parameter $M^2$ and the continuum threshold $s_0$.
In the QCDSR approach, auxiliary parameters are introduced as mathematical tools and, in principle, should not influence the extracted physical quantities. In practice, however, exact independence cannot be achieved. The working intervals are therefore determined following the standard prescriptions of the method. The selection criteria include minimal sensitivity of the TFFs to these parameters, maximizing the pole contribution (PC),  and ensuring the convergence of the operator product expansion (OPE). To satisfy the requirements of PC and OPE, the following two constraints are imposed \cite{Dehghan:2025eov}:
\allowdisplaybreaks
\begin{equation}
\begin{split}
	&	\text{PC}(Q^2) = \frac{\Pi_{i}^{\text{QCD}}(s_0, M^2, Q^2)}{\Pi_{i}^{\text{QCD}}(\infty, M^2, Q^2)} \geqslant 0.5,
	\\
	&\text{R} (M^2, Q^2) = \frac{\Pi_{i}^{\text{QCD, Dim6}}(s_0, M^2, Q^2)}{\Pi_{i}^{\text{QCD}}(\infty, M^2, Q^2)} \leqslant 0.08,
\end{split}
\end{equation}
where $i$ represent the specific Lorentz structures.
The upper bound of the Borel parameter $M^2$ is fixed by the PC criterion, which requires that at least half of the total QCD contribution originate from the ground-state pole.
The lower bound is set by the OPE convergence criterion, which restricts the contribution of the highest-dimensional term retained in the expansion to at most $8\%$ of the total QCD result.
From our analysis, the working intervals for $s_0$ and $M^2$ are determined as follows:
	\begin{equation}
		\begin{split}
			&s_0\in\left[ 2.9 ~\text{GeV}^2, 3.3 ~\text{GeV}^2\right], \\
			&M^2\in\left[3 ~\text{GeV}^2, 4 ~\text{GeV}^2\right].
		\end{split}
	\end{equation}
	We investigate the behavior of the TFFs as functions of $M^2$ and $Q^2$ in Figs.~\ref{Msqisoscalar}-\ref{Qsqisovector}.
	The TFFs’ behavior as a function of the Borel mass $M^2$ is analyzed at $Q^2 = 1.0$ GeV$^2$ for three values of $s_0$, and is plotted in Figs.~\ref{Msqisoscalar} and \ref{Msqisovector} for the isoscalar and isovector cases, respectively.
	The results indicate that the TFFs remain stable under variations of the Borel parameter within the working region, confirming the consistency of the QCDSR predictions.
	The behavior of the TFFs as functions of $Q^2$ is shown in Figs.~\ref{Qsqisoscalar} and \ref{Qsqisovector} for the isoscalar and isovector cases, respectively, at a fixed Borel mass parameter $M^2=3.5\ \text{GeV}^2$ and three values of the continuum threshold, $s_0 = 2.9, 3.1$ and $3.3\ \text{GeV}^2$.
	The results show that the TFFs decrease smoothly with increasing $Q^2$.
	The analysis of the QCDSR results indicates that the TFFs of the $\Delta^+$ baryon are well reproduced by the generalized $\mathbf{p}$-pole fit formula \cite{Dehghan:2023ytx},
	\begin{equation}
		{\cal F}(Q^2)=\frac{{\cal F}(0)}{\left(1+\frac{Q^2}{m_\mathbf{p}^2} \right)^\mathbf{p} },
		\label{dipole fit}
	\end{equation}
	where $m_\mathbf{p}$ represents an effective mass scale with dimension of energy (GeV), and the  $\mathbf{p}$ and ${\cal F}(0)$ are dimensionless parameters.
	The $\mathbf{p}$-pole fit functions of the isoscalar and isovector TFFs of the $\Delta^+$ baryon decrease and approach vanishing values as $Q^2$ reaches about $10\ \text{GeV}^2$, as shown in Figs.~\ref{Qsqisoscalar} and \ref{Qsqisovector}.
	The $\mathbf{p}$-pole fit parameters of the isoscalar and isovector TFFs, shown in Figs.~\ref{Qsqisoscalar} and \ref{Qsqisovector}, are summarized in Tables~\ref{fitparameters-isoS} and~\ref{fitparameters-isoV}, respectively. The fits are obtained at the mean values of $s_0$ and $M^2$, with uncertainties due to auxiliary parameters variations and systematic effects of the QCDSR approach.

	\begin{table}[!htb]
		\centering
		\begin{minipage}{0.48\textwidth}
			\centering
			\renewcommand{\arraystretch}{1.4} % increases row height
			\setlength{\tabcolsep}{10pt}      % increases column padding
			\begin{tabular}{|c|c|c|c|}
				\hline
				TFF &${\cal F}(0)$ &  $m_\mathbf{p}$ &  $\mathbf{p}$\\
				\hline\hline
				 $F^T_{1,0}(Q^2)$  &  $3.53 \pm 0.52  $  & $ 1.33 \pm 0.01$ & $1.22 \pm 0.06$   \\
				\hline
					$F^T_{1,1}(Q^2)$   &  $0.45 \pm 0.04$  & $1.68 \pm 0.04$ & $1.90 \pm 0.02$   \\
				\hline
				 $F^T_{2,0}(Q^2)$ &  $0.00 \pm 0.00$ & $1.76 \pm 0.06$ & $0.94 \pm 0.00$  \\
				\hline
				 $F^T_{3,0}(Q^2)$ &  $-0.04 \pm 0.00$ & $0.59 \pm 0.01$ & $1.01 \pm 0.01$ \\
				\hline
				 $F^T_{3,1}(Q^2)$ & $-2.11 \pm 0.21$ & $0.65 \pm 0.01$  & $1.55 \pm 0.02$  \\
				\hline
					$F^T_{4,0}(Q^2)$  &  $-0.09 \pm 0.01$ & $1.58 \pm 0.04$ & $2.10 \pm 0.01$  \\
				\hline
					$F^T_{5,0}(Q^2)$  &  $-0.00 \pm 0.00$  & $4.54 \pm 0.51$  &  $3.66 \pm 0.55$  \\
				\hline
					$F^T_{5,1}(Q^2)$ &  $1.11 \pm 0.14$  &$0.65 \pm 0.01$  &  $1.76 \pm 0.02$  \\
				\hline
					$F^T_{6,0}(Q^2)$& $0.01 \pm 0.00$ &$1.19 \pm 0.03$  & $0.94 \pm 0.01$ \\
				\hline
					$F^T_{7,0}(Q^2)$ &  $-0.04 \pm 0.02$ &$0.61 \pm 0.00$  & $0.81 \pm 0.01$  \\
				\hline
			\end{tabular}
			\caption{The numerical values of \textbf{p}-pole fit parameters of the isoscalar TFFs in Fig.~\ref{Qsqisoscalar} at the mean values of the continuum threshold $s_0$ and Borel mass parameter $M^2$.}
			\label{fitparameters-isoS}
		\end{minipage}
			\hfill
		\begin{minipage}{0.48\textwidth}
		\centering
		\renewcommand{\arraystretch}{1.4} % increases row height
		\setlength{\tabcolsep}{10pt}      % increases column padding
			\begin{tabular}{|c|c|c|c|}
				\hline
				TFF &  ${\cal F}(0)$ & $m_\mathbf{p}$ &  $\mathbf{p}$\\
				\hline\hline
				$F^T_{1,0}(Q^2)$ &  $ 0.74 \pm 0.10 $  & $ 1.17 \pm 0.03$ & $0.99 \pm 0.01$ \\
				\hline
					$F^T_{1,1}(Q^2)$ &  $0.09 \pm 0.01$ & $1.68 \pm 0.04$ & $1.90 \pm 0.02$\\
				\hline
				$F^T_{2,0}(Q^2)$ &  $-3.31 \pm 0.27$ & $1.17 \pm 0.03$ & $0.99 \pm 0.01$ \\
				\hline
				$F^T_{3,0}(Q^2)$  &  $-2.26 \pm 0.22$  & $0.66 \pm 0.01$ &$1.36 \pm 0.02$  \\
				\hline
				$F^T_{3,1}(Q^2)$  &  $2.14 \pm 0.21$  & $0.64 \pm 0.01$ & $1.49 \pm 0.02$ \\
				\hline
				$F^T_{4,0}(Q^2)$ &  $-0.98 \pm 0.09$  & $0.66 \pm 0.01$ & $1.78 \pm 0.02$ \\
				\hline
				$F^T_{5,0}(Q^2)$ & $-2.26 \pm 0.22$ & $0.66 \pm 0.01$& $1.36 \pm 0.02$ \\
				\hline
				$F^T_{5,1}(Q^2)$ & $-1.90 \pm 0.30$ &$0.65 \pm 0.01$ & $1.75 \pm 0.02$   \\
				\hline
				$F^T_{6,0}(Q^2)$ & $-0.01 \pm 0.00$ & $0.53 \pm 0.05$ & $1.09 \pm 0.01$  \\
				\hline
				$F^T_{7,0}(Q^2)$ & $-0.08 \pm 0.01$ &$0.59 \pm 0.00$ & $0.80 \pm 0.15$  \\
				\hline
			\end{tabular}
			\caption{The numerical values of \textbf{p}-pole fit parameters of the isovector TFFs in Fig.~\ref{Qsqisovector} at the mean values of the continuum threshold $s_0$ and Borel mass parameter $M^2$.}
			\label{fitparameters-isoV}
		\end{minipage}
	\end{table}
	
As shown in Tables \ref{fitparameters-isoS} and \ref{fitparameters-isoV}, the fit parameters obtained for the isoscalar and isovector TFFs are different. This difference arises from the distinct flavor structures of the corresponding tensor currents: the isoscalar current couples to the sum of the
$u$- and $d$-quark contributions, while the isovector current couples to their difference. As a result, these currents probe distinct combinations of quark tensor densities, which leads to the observed numerical differences in the TFFs.
Understanding the distinction between the isoscalar and isovector TFFs is important, as it provides information about how the tensor charge is distributed among different quark flavors and how isospin symmetry is reflected in the internal dynamics of the $\Delta^+$ baryon.
 In Ref.~\cite{Fu:2024kfx}, the quark tensor charge is extracted in the forward limit \textquotedblleft $q=0$\textquotedblright.
 From Eq.~\eqref{martix-element} and in the forward limit, 
 \allowdisplaybreaks
 \begin{align}
	\left\langle \Delta^+(p,s)\left|  \bar{\psi}(0)i\sigma_{\mu\nu}\psi(0)\right| \Delta^+(p,s)\right\rangle=\bar{u}^\alpha(p,s)
\Big[i\sigma_{\mu\nu} g_{\alpha\beta}F^T_{1,0}(0) 
+ig_{\alpha [\mu}\sigma _{\nu]\beta}F_{2,0}^T(0)
\Big]u^\beta(p,s).
 \end{align}
 Using the Rarita–Schwinger constraints in Eq. \eqref{RS} and the antisymmetry of $\sigma_{\mu\nu}$, we get
  \allowdisplaybreaks
 \begin{align}
 	\left\langle \Delta^+(p,s)\left|  \bar{\psi}(0)i\sigma_{\mu\nu}\psi(0)\right| \Delta^+(p,s)\right\rangle=\bar{u}^\alpha(p,s) i\sigma_{\mu\nu} {u}_\alpha(p,s)
 	\Big(F^T_{1,0}(0) -F_{2,0}^T(0)\Big),
 \end{align}
where the nonzero combination of TFFs, $F_{1,0}^T(0)-F_{2,0}^T(0)$ corresponds to the quark tensor charge, which we denote by $\delta\psi$. This result is in agreement with \cite{Fu:2024kfx}.  Using the values extracted in Tables \ref{fitparameters-isoS} and \ref{fitparameters-isoV}, the quark tensor charge can be calculated for both the isoscalar and isovector cases.
\allowdisplaybreaks
\begin{align}
&\delta\psi^{\text{isoscalar}}=3.53\pm0.52,\nonumber\\
&\delta\psi^{\text{isovector}}=4.05\pm0.29.
\end{align}
Moreover, small deviations may also arise from flavor symmetry breaking effects, such as the small mass difference between the $u$ and $d$ quarks or differences in their condensates.
Such effects can change the contribution of each tensor current, leading to a larger separation between the isoscalar and isovector TFFs.

As mentioned, the tensor charge provides essential information about the internal spin structure of the hadron and reflects the contribution of quark transversity to its intrinsic properties. It characterizes the response of the hadron to tensor interactions and provides a fundamental measure of the correlations among quark spin degrees of freedom inside the hadron.
Beyond the Standard Model, the tensor charge is interpreted as the intrinsic electric dipole moment carried by the corresponding quark \cite{Fu:2024kfx, Pospelov:2005pr, Wang:2018kto}.
The other TFFs would provide insight into various aspects of the physics of the tensor current interaction with spin-${3}/{2}$ particles, including spin distributions, electromagnetic characteristics, and possible indications of their geometric shape. In the future, different physical observables may be constructed in which these TFFs would serve as the fundamental building blocks. 

	\section{Summary and conclusion}\label{conclusion}
Because hadrons interact through different types of currents, each interaction is characterized by a distinct set of FFs. Determining these FFs yields valuable information about various physical properties.
Although some experimental data on hadronic interactions are available, there are still differences between theory and experiment. These differences show that hadron structure is very complex and still not completely explained. To address this, we need more precise theoretical studies and improved experiments in the future. 
For spin-$\frac{3}{2}$ particles, there is still very little information available from both the theory and experiment. This lack of data means that we do not yet fully understand their structure, interactions, and properties. While some theoretical studies have investigated their electromagnetic and gravitational interactions, these offer give a partial picture. To gain a clearer understanding of these particles, it is necessary to study their interactions with other types of currents, such as the tensor current.

TFFs of decuplet baryons are important because they help us understand the spin structure of these baryons and show detailed properties, such as the distribution of tensor charge. These FFs, obtained from the matrix elements of the tensor current between hadron states, provide a useful way to study the shape and internal dynamics of baryons and play a key role in understanding the nonperturbative features of QCD.
With recent progress in studying the interactions of these currents, especially tensor and gravitational currents with hadrons, more precise experimental data are expected  soon. These new data will not only help clarify the properties of spin-$3/2$ particles but also significantly improve our understanding of the structure and dynamics of nucleons.
This data serve as important input for experiments at leading research centers around the world, such as the Large Hadron Collider (LHC), JLab, Mainz, LAPP, and other advanced laboratories. These centers can provide valuable information on the interactions of various currents with nucleons and spin-$3/2$ particles.

In this work, we  studied the $\Delta^+ \to \Delta^+$ transition induced by the tensor current.
The tensor matrix element for decuplet baryons has also been formulated in terms of seven TFFs through the first moments of the transversity GPDs in Ref.~\cite{Fu:2024kfx}.  In present study, we have derived the tensor matrix element for the transition $\frac{3}{2}^+\to\frac{3}{2}^+$ using a completely independent approach, expressing it in terms of ten TFFs. 
All the structures are independent and respect time-reversal, Hermiticity, and parity symmetries.
These two approaches have both similarities and differences.  Six of the TFFs are in complete agreement between the two methods. 
However, in our formulation we employed the Lorentz structures that explicitly satisfy the discrete symmetries.
The additional TFFs in our approach guarantee full consistency between the physical and QCD sides structures while respecting all discrete symmetries.
This decomposition provides the basis for calculating the physical side of the  QCDSR framework. On the other hand, the QCD side of the correlation function was calculated separately for the isoscalar and isovector tensor currents.
Then, the coefficients of the corresponding Lorentz structures were matched between the physical and QCD representations. Using this procedure, we derived the ten TFFs and numerically evaluated them for the $\Delta^+$ baryon in $ Q^2\in \left[0 , 10 \right] $ GeV$^2$ for both the isoscalar and isovector cases, within the three-point QCDSR framework. The QCDSR  method is a fully relativistic approach that accounts for various hadronic properties and quantum numbers, including spin, making it one of the leading nonperturbative techniques. We found that the $Q^2$ dependence of the $\Delta^+$’s TFFs is well described by a \textbf{p}-pole fit function, and we also reported the results for the TFFs at $Q^2=0$.
The observed differences between the isoscalar and isovector TFFs reflect the fact that the corresponding tensor currents couple to different components of the baryon’s internal quark structure. The isoscalar TFFs represent the total tensor response of both light quarks, whereas the isovector TFFs isolate the relative contributions of the up and down quarks. Consequently, even in the isospin-symmetric limit, their magnitudes and $Q^2$-dependence are expected to differ.
The results presented in this work can serve as a useful reference for ongoing and future experimental programs.
In particular, facilities such as JLab have indicated  the study of spin-$\frac{3}{2}$ baryons and the determination of their GPDs as key objectives. These studies play an important role in calculating and identifying the various TFFs of these particles, providing detailed information about their internal structure and the dynamics of their interactions.

\appendix
\renewcommand{\thesection}{\Alph{section}}
\renewcommand{\thesubsection}{\thesection. \arabic{subsection}}
	\section{The matrix element of the tensor current between decuplet baryon states} \label{appA}
	In this section, we extract the tensor matrix element for the $\frac{3}{2}^+\to\frac{3}{2}^+$  transition and start by writing the general form of the matrix element as
	\begin{equation}
		\left\langle  B_f(p^\prime,s^\prime)\left|  \bar{\psi}(0)i\sigma_{\mu\nu}\psi(0)\right| B_i(p,s)\right\rangle=\bar{u}^\alpha(p^\prime,s^\prime)\mathcal{O}_{\alpha\beta\mu\nu}u^\beta(p,s),
		\label{general}
	\end{equation}
	where 	$\mathcal{O}_{\alpha\beta\mu\nu}$ is the transition operator, constructed from all possible Lorentz structures and forming a rank-$4$ tensor in the indices $(\mu\nu)$ and $(\alpha\beta)$.
	Its building blocks consist of the metric tensor, gamma matrices, momentum transfer, and total momentum, together with the corresponding TFFs. The structures are required to be antisymmetric under
	$\mu\leftrightarrow \nu$.
	
	We first construct all possible Lorentz structures, giving a total of $32$ terms. 
	To isolate the independent structures, we use the on-shell identities, written as \cite{Cotogno:2019vjb},
	\allowdisplaybreaks
	\begin{align}
		&i\sigma_{\nu\beta}\doteq g_{\nu\beta},\nonumber
		\\
		&\frac{\slashed \mtotal}{2m}\doteq \textbf{1},\quad\slashed{q}\doteq m_i-m_f,\nonumber
		\\
		&\gamma_\mu\doteq\frac{\mtotal_\mu}{2m}-\frac{i\sigma_{\mu\nu}q^\nu}{2m},\nonumber
		\\
		&	2m q_{[\alpha}g_{\beta][\mu}\mtotal_{\nu]}\doteq\mtotal_{[\alpha}q_{\beta]}\mtotal_{[\mu}\gamma_{\nu]}+\mtotal^2 q_{[\alpha}g_{\beta][\mu}\gamma_{\nu]},
		\label{identities}
	\end{align}
	where $m={\left( m_i+m_f\right) }/{2}$, with $m_i$ and $m_f$ representing the masses of the decuplet baryons in the initial and final states, respectively. A contraction with a four-vector is indicated by substituting the corresponding Lorentz index with that vector,  $i\sigma_{\nu q}=i\sigma_{\nu \rho}q^\rho$.
	Furthermore, we use the product of two Dirac matrix $\gamma_\beta\gamma_\mu=2g_{\beta\mu}-\gamma_\mu\gamma_\beta$, 
	together with the Rarita–Schwinger constraints,
	\begin{equation}
		\begin{split}
			&\gamma_{\beta}u^\beta(p,s)=0,\quad p_\beta u^\beta(p,s)=0.
		\end{split}
		\label{RS}
	\end{equation}
	Consequently, the structures were reduced to 12 after removing those that were not independent or could be expressed in terms of others.
	All tensor structures are required to satisfy Hermiticity, which leads to the following constraint \cite{Cotogno:2019vjb},
	\\
	\allowdisplaybreaks
	\begin{align}
		& 
		\left\langle B_f\left( p',s'\right)  \left|  \bar{\psi}(0)i\sigma_{\mu\nu}\psi(0)\right| B_i\left( p,s\right)\right\rangle 
		=-\left\langle B_i\left( p,s\right) \left| \bar{\psi}(0)i\sigma_{\mu\nu}\psi(0)\right| B_f\left( p',s'\right) \right\rangle^*.
		\label{HTP}
	\end{align}
	On the other hand, imposing parity and time-reversal symmetries leads to the following combined constraint \cite{Cotogno:2019vjb},
	 \allowdisplaybreaks
	 \begin{align}
	 	& 
	 	\mathcal{O}_{\alpha\beta\mu\nu}(\mtotal,q)
	 	=-\gamma_0\left[ \mathcal{O}_{\beta\alpha\mu\nu}(\mtotal,-q)\right] ^\dagger\gamma_0.
	 	\label{HTP1}
	 \end{align}
	 Here,  the explicit dependence of the operator $\mathcal{O}_{\alpha\beta\mu\nu}$ on the metric tensor and the gamma matrices is suppressed for simplicity.
	The minus sign appearing on the right-hand side, outside the brackets in Eqs.~\eqref{HTP} and \eqref{HTP1}, arises from the antisymmetric nature of the tensor current in the matrix element.
	Any structure that does not satisfy the conditions in Eqs.~\eqref{HTP} and \eqref{HTP1} is discarded, thereby reducing the number of independent structures to ten. Finally, we obtain the matrix element of the tensor current for the transition $\frac{3}{2}^+\to\frac{3}{2}^+$ as
	\begin{equation}
		\begin{split}
			\left\langle  B_f(p^\prime,s^\prime)\left|  \bar{\psi}(0)i\sigma_{\mu\nu}\psi(0)\right| B_i(p,s)\right\rangle&=\bar{u}^\alpha(p^\prime,s^\prime)
			\Bigg[i\sigma_{\mu\nu} \bigg( g_{\alpha\beta}F^T_{1,0}(Q^2)+\frac{q_{\alpha}q_{\beta}}{m^2}F^T_{1,1}(Q^2)\bigg) 
			+ig_{\alpha [\mu}\sigma _{\nu]\beta}F_{2,0}^T(Q^2)
			\\&+\frac{\gamma_{[\mu}q_{\nu]}}{m}\bigg(g_{\alpha\beta}F_{3,0}^T(Q^2)+\frac{q_{\alpha}q_{\beta}}{m^2}F^T_{3,1}(Q^2) \bigg) 
			+\frac{\gamma_{[\mu}\mtotal_{\nu]}\mtotal_{[\alpha}q_{\beta]}}{m^3} F^T_{4,0}(Q^2)\\
			&+\frac{\mtotal_{[\mu}q_{\nu]}}{m^2}\bigg(g_{\alpha\beta}F_{5,0}^T(Q^2)+\frac{q_{\alpha}q_{\beta}}{m^2}F^T_{5,1}(Q^2) \bigg)
			+\frac{g_{\alpha[\mu}\gamma_{\nu]}q_\beta-g_{\beta[\mu}\gamma_{\nu]}q_\alpha}{m}F^T_{6,0}(Q^2)\\&+\frac{g_{\alpha[\mu}q_{\nu]}\mtotal_\beta-g_{\beta[\mu}q_{\nu]}\mtotal_\alpha}{m^2}F^T_{7,0}(Q^2)
			\Bigg]u^\beta(p,s).
			\label{final-martix}
		\end{split}
	\end{equation}
		As a result, we obtained the tensor matrix element for the transition $\frac{3}{2}^+\to\frac{3}{2}^+$ in terms of ten TFFs, all of which are independent and fully respect the discrete symmetries of T-invariance, Hermiticity, and parity.
		
	On the other hand, the semileptonic matrix element for the transition $\frac{3}{2}^+\to\frac{3}{2}^+$  is \cite{Faessler:2009xn}, 
	\begin{equation}
		\begin{split}
			\left\langle  B_f(p^\prime,s^\prime)\left| J_\mu^V+J_\mu^A\right| B_i(p,s)\right\rangle&=	\left\langle  B_f(p^\prime,s^\prime)\left|  \bar{\psi}(0)[i\sigma_{\mu\nu}\dfrac{q^\nu}{m}(1+\gamma_5)]\psi(0)\right| B_i(p,s)\right\rangle\\
			&=\bar{u}^\alpha(p^\prime,s^\prime)\Bigg[g_{\alpha\beta}\bigg(\gamma_\mu F^V_1(Q^2)+i\sigma_{\mu\nu}\frac{q^\nu}{m}F^V_2(Q^2)
			+\frac{q_\mu}{m}F^V_3(Q^2)\bigg) \\&
			+\frac{q_{\alpha}q_{\beta}}{m^2}\bigg(\gamma_\mu F^V_4(Q^2) +i\sigma_{\mu\nu}\frac{q^\nu}{m}F^V_5(Q^2)
			+\frac{q_\mu}{m}F^V_6(Q^2)\bigg) 
			+\frac{g_{\alpha\mu}q_\beta-g_{\beta\mu}q_\alpha}{m}F^V_7(Q^2)\Bigg]u^\beta(p,s)\\
			&
			+\bar{u}^\alpha(p^\prime,s^\prime)\Bigg[g_{\alpha\beta}\bigg(\gamma_\mu F^A_1(Q^2)
			+i\sigma_{\mu\nu}\frac{q^\nu}{m}F^A_2(Q^2)
			+\frac{q_\mu}{m}F^A_3(Q^2)\bigg) \\&
			+\frac{q_{\alpha}q_{\beta}}{m^2}\bigg(\gamma_\mu F^A_4(Q^2) +i\sigma_{\mu\nu}\frac{q^\nu}{m}F^A_5(Q^2)
			+\frac{q_\mu}{m}F^A_6(Q^2)\bigg)
			+\frac{g_{\alpha\mu}q_\beta-g_{\beta\mu}q_\alpha}{m}F^A_7(Q^2)\Bigg]\gamma_5u^\beta(p,s).
			\label{semi2}
		\end{split}
	\end{equation}
	To ensure the validity of the derived tensor matrix element, we multiply both sides of Eq.~\eqref{final-martix} by $q^\nu(1+\gamma_5)/m$, yielding Eq.~\eqref{semi2}.
	The semileptonic FFs are then obtained as linear combinations of the TFFs,
	\allowdisplaybreaks
	\begin{align}
		&F^V_1(Q^2)=- \frac{Q^2}{m^2}F^T_{3,0}(Q^2)-\frac{2Q^2}{m^2}F_{5,0}^T(Q^2),
		\nonumber	\\
		&F^V_2(Q^2)= F^T_{1,0}(Q^2)-\frac{Q^2}{m^2}F_{5,0}^T(Q^2),
		\nonumber	\\
		&F^V_3(Q^2)=-\frac{2\left( m_i-m_f\right) }{m}F_{5,0}^T(Q^2)-\frac{\left( m_i-m_f\right)}{m} F_{3,0}^T(Q^2),
		\nonumber\\
		&F^V_4(Q^2)=- \frac{Q^2}{m^2}F^T_{3,1}(Q^2)-\frac{2Q^2}{m}F^T_{5,1}(Q^2),
		\nonumber\\
		&F^V_5(Q^2)=F^T_{1,1}(Q^2) -\frac{Q^2}{m^2}F^T_{5,1}(Q^2)-\frac{2(m_i-m_f)}{m}F^T_{4,0}(Q^2),
		\nonumber	\\
		&F^V_6(Q^2)=-\frac{\left( m_i-m_f\right)}{m}F^T_{3,1}(Q^2) -\frac{2\left( m_i-m_f\right) }{m}F^T_{5,1}(Q^2)+2F^T_{7,0}(Q^2),
		\nonumber\\
		&F^V_7(Q^2)=-F^T_{2,0}(Q^2)+\frac{\left( m_i-m_f\right) }{m}F^T_{6,0}(Q^2)-\frac{Q^2}{m^2} F^T_{7,0},
		\nonumber	\\
		&F^A_1(Q^2)=- \frac{Q^2}{m^2}F^T_{3,0}(Q^2)-\frac{2Q^2}{m^2}F_{5,0}^T(Q^2),
		\nonumber	\\
		&F^A_2(Q^2)= F^T_{1,0}(Q^2)-\frac{Q^2}{m^2}F_{5,0}^T(Q^2),
		\nonumber	\\
		&F^A_3(Q^2)=-\frac{2\left( m_i-m_f\right) }{m}F_{5,0}^T(Q^2)+2F_{3,0}^T(Q^2),
		\nonumber	\\
		&F^A_4(Q^2)=- \frac{Q^2}{m^2}F^T_{3,1}(Q^2)-\frac{2Q^2}{m}F^T_{5,1}(Q^2)-4\left( m_i+m_f\right)F^T_{4,0}(Q^2),
		\nonumber	\\
		&F^A_5(Q^2)= F^T_{1,1}(Q^2) -\frac{Q^2}{m^2}F^T_{5,1}(Q^2)-4F^T_{4,0}(Q^2),
		\nonumber	\\
		&F^A_6(Q^2)= 2F^T_{3,1}(Q^2) -\frac{2\left( m_i-m_f\right) }{m}F^T_{5,1}(Q^2)+2F^T_{7,0}(Q^2),
		\nonumber	\\
		&F^A_7(Q^2)=-F^T_{2,0}(Q^2)-2F^T_{6,0}(Q^2)-\frac{Q^2}{m^2} F^T_{7,0}.
		\label{semi}
	\end{align}
	This establishes a useful connection between the TFFs and the semileptonic FFs
	
	 In this step, we review the approach of Ref.~\cite{Fu:2024kfx} for the tensor matrix element and discuss both the points of agreement and the differences between the results obtained from the two approaches.
	
	GPDs provide a more comprehensive description of hadron structure in terms of their fundamental degrees of freedom than conventional parton distributions and FFs.
Deep inelastic scattering probes the leading-twist parton distributions, such as polarized, unpolarized, and transversity, which are further generalized by GPDs to describe the corresponding parton distributions within hadrons.
The matrix elements of the transverse nonlocal quark–quark correlator determine the quark transversity GPDs \cite{Fu:2024kfx}:
	\allowdisplaybreaks
	\begin{align}
		T_{\tau}^{h' h} = -\bar{u}^{\alpha} (p',h')
		\mathcal{H}_{T,\tau, \alpha \beta}(x,\xi,Q^2) u^\beta(p,h),
	\end{align}
	where $\tau$ labels the transverse indices, and $h (h')$ corresponds to the helicity of the initial (final) state. The variables $x$ and $\xi$ denote the longitudinal momentum fraction and the skewness parameter, respectively.
	The tensor function $\mathcal{H}_{T, \tau, \alpha\beta}(x,\xi,Q^2)$ is decomposed into sixteen independent tensor structures as coefficients of sixteen GPDs \cite{Fu:2024kfx}.
The first moments of the transversity GPDs define the TFFs, with only the terms of $\mathcal{H}_{T, \tau, \alpha\beta}(x,\xi,Q^2)$ yielding non-vanishing integrals contributing directly to them.

Among the sixteen terms of the tensor function, only seven yield nonzero contributions upon integration, while the integrals of the remaining terms vanish. Consequently, this approach produces seven TFFs, and the tensor matrix element is expressed as \cite{Fu:2024kfx},
\allowdisplaybreaks
\begin{align}
	\left\langle p', h' \left| \bar{\psi} (0)
	i \sigma_{\mu \nu} \psi (0) \right|p, h\right\rangle&=-2\ \bar{u}^{\alpha} (p',h')\Bigg[g_{\alpha\beta}\bigg(G^{ T}_1(Q^2)
	i \sigma_{\mu \nu} + G^{ T}_5(Q^2) \frac{\gamma_{[ \mu}
		\tilde{q}_{ \nu ]}}{m}
	+ G^{ T}_7(Q^2) \frac{\tilde{\mtotal}_{[ \mu } \tilde{q}_{\nu ]}}{m^2} \bigg)+ G^{ T}_{2}(Q^2) g_{\mu [\alpha}\ g_{\beta ] \nu}\nonumber
	\\
	&+ \frac{\tilde{\mtotal}_\alpha\tilde{\mtotal}_{\beta} }{m^2}\bigg( G^{ T}_6(Q^2)
	\frac{\gamma_{[ \mu} \tilde{q}_{ \nu ]}}{m} + G^{ T}_8(Q^2)
	\frac{\tilde{\mtotal}_{[ \mu} \tilde{q}_{ \nu ]}}{m^2} \bigg)
	+ G^{ T}_{12}(Q^2) \frac{\tilde{\mtotal}_{[\alpha}\ g_{\beta]
			[\nu } \tilde{\mtotal}_{\mu ]}}{m^2}
	\Bigg] u^\beta(p,h),
	\label{TFF GPD}
\end{align}
where $G^{T}_\tau (Q^2)$ with $\tau=1, 2, 5-8, 12$ denotes the TFFs, which are independent of the skewness parameter $\xi$.
The tensor matrix elements in Eqs. \eqref{final-martix} and \eqref{TFF GPD} are expressed using different conventions for the total momentum and momentum transfer, with $\tilde{\mtotal}_\mu=\mtotal_\mu/2$ and $\tilde{q}_\mu=-q_\mu$.
	 By comparing the tensor matrix elements in Eqs.~\eqref{final-martix} and \eqref{TFF GPD}, we observe
	\allowdisplaybreaks
	\begin{align}
&F^T_{1,0}(Q^2)=-2G_1^T(Q^2),\nonumber\\
&F^T_{3,0}(Q^2)=2G_5^T(Q^2),\nonumber\\
&F^T_{3,1}(Q^2)=-\frac{1}{2}G_6^T(Q^2),\nonumber\\
&F^T_{5,0}(Q^2)=G_7^T(Q^2),\nonumber\\
&F^T_{5,1}(Q^2)=-\frac{1}{4}G_8^T(Q^2).
	\end{align}
It should be noted that in Eq. \eqref{TFF GPD} the term $\mtotal_\alpha\mtotal_{\beta} $ appears, whereas in Eq. \eqref{final-martix} the corresponding structure is $q_\alpha q_{\beta} $. Using the definitions of the momentum transfer and the total momentum,
\allowdisplaybreaks
\begin{align}
	&p^\prime_\beta= p_\beta-q_\beta\Rightarrow \ {\mtotal_\beta=2p_\beta-q_\beta}, \nonumber\\
	&p_\alpha=p^\prime_\alpha+q_\alpha,\ \Rightarrow\ {\mtotal_\alpha=2p^\prime_\alpha+q_\alpha},
	\label{relation}
\end{align}
together with the Rarita–Schwinger constraints, one can readily show that,
	\begin{equation}
	\begin{split}
		\bar{u}^\alpha(p^\prime,s^\prime)\big[\mtotal_\alpha \mtotal_\beta\big]u^\beta(p,s)&=	\bar{u}^\alpha(p^\prime,s^\prime)\big[4p^\prime_\alpha p_\beta-2p^\prime_\alpha q_\beta +2p_\beta q_\alpha-q_\alpha q_\beta\big]u^\beta(p,s)=-\bar{u}^\alpha(p^\prime,s^\prime)\big[q_\alpha q_\beta\big]u^\beta(p,s).
	\end{split}
\end{equation}
	Using the on-shell identities in Eq.~\eqref{identities},
	\allowdisplaybreaks
	\begin{equation}
	\begin{split}
	& -\dfrac{2}{m^2}\tilde{\mtotal}_{[\alpha}\ g_{\beta]
		[\nu } \tilde{\mtotal}_{\mu ]}\doteq\dfrac{1}{m}\tilde{q}_{[\alpha}\ g_{\beta]
		[\mu }{\gamma}_{\nu ]}+\dfrac{1}{m^3}\tilde{\mtotal}_{[\mu}\gamma_{\nu]}\tilde{\mtotal}_{[\alpha}\tilde{q}_{\beta]},
		\\
		&g_{\mu [\alpha}\ g_{\beta ] \nu}\doteq ig_{\alpha [\mu}\sigma _{\nu]\beta},
\end{split}
	\end{equation}
	as a result, 
		\allowdisplaybreaks
		\begin{equation}
	\begin{split}
		&F^T_{6,0}(Q^2)+\frac{1}{4}F^T_{4,0}(Q^2)=G_{12}^T(Q^2),
		\\
		&F^T_{2,0}(Q^2)=-2G_2^T(Q^2),	
	\end{split}
	\end{equation}
	 However, according to the constraints in Eq.~\eqref{HTP}, the TFFs $G_2^T(Q^2)$ and $G_{12}^T(Q^2)$ appearing in Eq.~\eqref{TFF GPD} do not satisfy Hermiticity, parity and T-invariance.
	Eq.~\eqref{final-martix} includes TFFs that respect to the mentioned symmetries and fully ensure consistency between the physical and QCD sides structures, which is not completely achieved in Eq.~\eqref{TFF GPD}.

	\section{Hadronic side results} \label{appB}
	In this appendix, we present the explicit forms of the functions $\Pi_i^{\text{Had}}(Q^2)$ in terms of the TFFs and other hadronic parameters.
\begin{equation}
	\begin{array}{l@{\hspace{2cm}}l@{\hspace{2cm}}l} % فاصله 2cm بین ستون‌ها
		\Pi_1^{\text{Had}}(Q^2)=F_{1,0}^T(Q^2), 
		&
		\Pi_{8}^{\text{Had}}(Q^2)=-\dfrac{2}{m_\Delta^3}\left(F^T_{3,1}(Q^2)-2 F^T_{4,0}(Q^2)+F^T_{5,1}(Q^2)\right) ,
		\\
		\\
		\Pi_2^{\text{Had}}(Q^2)=-F_{2,0}^T(Q^2),
		&
		\Pi_{9}^{\text{Had}}(Q^2)=\dfrac{2}{m_\Delta^2}\left(F^T_{3,1}(Q^2)-2 F^T_{4,0}(Q^2)-F^T_{5,1}(Q^2)\right),
		\\
		\\
		\Pi_3^{\text{Had}}(Q^2)=-\dfrac{1}{m_\Delta^2}F^T_{1,1}(Q^2),
		&
		\Pi_{10}^{\text{Had}}(Q^2)=\dfrac{1}{m_\Delta^2}\left(F^T_{3,1}(Q^2)+2 F^T_{4,0}(Q^2)-2F^T_{1,1}(Q^2)\right),
		\\
		\\
		\Pi_4^{\text{Had}}(Q^2)=-\dfrac{1}{m_\Delta}F^T_{3,0}(Q^2),
		&
		\Pi_{11}^{\text{Had}}(Q^2)=\dfrac{1}{m_\Delta^2}\left(F^T_{3,1}(Q^2)+2F^T_{4,0}(Q^2) \right) , 
		\\
		\\
		\Pi_{5}^{\text{Had}}(Q^2)=\dfrac{2}{m_\Delta^2}F^T_{5,0}(Q^2),
		&
		\Pi_{12}^{\text{Had}}(Q^2)=\dfrac{1}{m_\Delta^2}\left(-F^T_{3,1}(Q^2)+2F^T_{4,0}(Q^2) \right),
		\\
		\\
		\Pi_6^{\text{Had}}(Q^2)=-\dfrac{1}{m_\Delta}F^T_{6,0}(Q^2), 
		& \Pi_{13}^{\text{Had}}(Q^2)=\dfrac{2}{m_\Delta}\left( F^T_{3,0}(Q^2)+F^T_{5,0}(Q^2)\right).
		\\
		\\
		\Pi_{7}^{\text{Had}}(Q^2)=\dfrac{1}{m_\Delta}F^T_{7,0}(Q^2),
	\end{array}
	\label{function}
\end{equation}
		\section{QCD side results} \label{appC}
	In this appendix, we present the three-point correlation function results for both the isovector and isoscalar tensor currents. 
	
	\vspace{0.3cm}
\underline{\textbf{\textit{{ Isoscalar case:}}}}
	\allowdisplaybreaks
\begin{align}
	\Pi^{\text{isoscalar}}_{\alpha\mu\nu\beta}(x,y)&=
	4\ \bigg[S_u^{c c^\prime}(y-x)\bigg] \text{Tr}\bigg[\ga S_u^{ae}(y-0) \s S_u^{e a^\prime}(0-x) \gb S_d^{\prime\ b b^\prime}(y-x)\bigg]
\nonumber	\\
	&-4\bigg[S_u^{c a^\prime}(y-x)\gb S_d^{\prime\ b b^\prime}(y-x)\ga S_u^{ae}(y-0) \s S_u^{e c^\prime}(0-x)\bigg]
\nonumber	\\
	&-4\bigg[S_u^{ce}(y-0)\s S_u^{e a^\prime}(0-x)\gb S_d^{\prime\ b b^\prime}(y-x)\ga S_u^{a c^\prime}(y-x)\bigg]
\nonumber	\\
	&+4\bigg[S_u^{ce}(y-0)\s S_u^{ec^\prime}(0-x)\bigg]\ \text{Tr}\bigg[\ga S_u^{aa^\prime}(y-x) \gb S_d^{\prime\ b b^\prime}(y-x)\bigg]
\nonumber	\\
	&-2\bigg[ S_u^{cb^\prime}(y-x)\gb S_u^{\prime\ ea^\prime}(0-x)\s S_u^{\prime\ ae}(y-0)\ga S_d^{bc^\prime}(y-x)\bigg]
\nonumber	\\
	&+2\bigg[ S_u^{ca^\prime}(y-x)\gb S_u^{\prime\ eb^\prime}(0-x)\s S_u^{\prime\ ae}(y-0)\ga S_d^{bc^\prime}(y-x)\bigg]
\nonumber	\\
	&+2\bigg[S_u^{ce}(y-0) \s S_u^{ea^\prime}(0-x) \gb S_u^{\prime\ ab^\prime}(y-x)\ga S_d^{bc^\prime}(y-x)\bigg]
	\nonumber\\
	&-2\bigg[S_u^{ce}(y-0) \s S_u^{eb^\prime}(0-x) \gb S_u^{\prime\ aa^\prime}(y-x)\ga S_d^{bc^\prime}(y-x)\bigg]
\nonumber	\\
	&-4\bigg[S_u^{ca^\prime}(y-x) \gb S_d^{\prime\ eb^\prime}(0-x)\s S_d^{\prime\ be}(y-0)\ga S_u^{ac^\prime}(y-x)\bigg]
\nonumber	\\
	&+4\bigg[S_u^{cc^\prime}(y-x)\bigg]\ \text{Tr}\bigg[\ga S_d^{be}(y-0) \s S_d^{eb^\prime}(0-x) \gb S_u^{\prime\ aa^\prime}(y-x)\bigg]
\nonumber	\\
	&+2\bigg[ S_u^{ca^\prime}(y-x)\gb S_u^{\prime\ ab^\prime}(y-x)\ga S_d^{be}(y-0) \s S_d^{ec^\prime}(0-x)\bigg]
\nonumber	\\
	&-2\bigg[S_u^{cb^\prime}(y-x)\gb S_u^{\prime\ aa^\prime}(y-x)\ga S_d^{be}(y-0)\s S_d^{ec^\prime}(0-x)\bigg]
\nonumber	\\
	&-2\bigg[S_d^{cb^\prime}(y-x)\gb S_u^{\prime\ ea^\prime}(0-x)\s S_u^{\prime\ ae}(y-0)\ga S_u^{bc^\prime}(y-x)\bigg]
\nonumber	\\
	&+2\bigg[S_d^{cb^\prime}(y-x)\gb S_u^{\prime\ ba^\prime}(y-x)\ga S_u^{ ae}(y-0)\s S_u^{ec^\prime}(0-x)\bigg]
\nonumber	\\
	&+2\bigg[S_d^{cb^\prime}(y-x)\gb S_u^{\prime\ ea^\prime}(0-x)\s S_u^{\prime\ be}(y-0)\ga S_u^{ac^\prime}(y-x)\bigg]
\nonumber	\\
	&-2\bigg[S_d^{cb^\prime}(y-x)\gb S_u^{\prime\ aa^\prime}(y-x)\ga S_u^{ be}(y-0)\s S_u^{ec^\prime}(0-x)\bigg]
\nonumber	\\
	&+\bigg[S_d^{cc^\prime}(y-x)\bigg]\ \text{Tr}\bigg[\ga S_u^{bb^\prime}(y-x)\gb S_u^{\prime\ ea^\prime}(0-x)\s S_u^{\prime\ ae}(y-0)\bigg]
\nonumber	\\
	&-\bigg[S_d^{cc^\prime}(y-x)\bigg]\ \text{Tr}\bigg[\ga S_u^{ba^\prime}(y-x)\gb S_u^{\prime\ eb^\prime}(0-x)\s S_u^{\prime\ ae}(y-0)\bigg]
\nonumber	\\
	&-\bigg[S_d^{cc^\prime}(y-x)\bigg]\ \text{Tr}\bigg[\ga S_u^{be}(y-0)\s S_u^{ ea^\prime}(0-x)\gb S_u^{\prime\ ab^\prime}(y-x)\bigg]
\nonumber	\\
	&+\bigg[S_d^{cc^\prime}(y-x)\bigg]\ \text{Tr}\bigg[\ga S_u^{be}(y-0)\s S_u^{ eb^\prime}(0-x)\gb S_u^{\prime\ aa^\prime}(y-x)\bigg]
\nonumber	\\
	&+2\bigg[S_d^{ce}(y-0)\s S_d^{eb^\prime}(0-x)\gb S_u^{\prime\ ba^\prime}(y-x)\ga S_u^{ac^\prime}(y-x)\bigg]
\nonumber	\\
	&-2\bigg[S_d^{ce}(y-0)\s S_d^{eb^\prime}(0-x)\gb S_u^{\prime\ aa^\prime}(y-x)\ga S_u^{bc^\prime}(y-x)\bigg]
\nonumber	\\
	&-\bigg[S_d^{ce}(y-0)\s S_d^{ec^\prime}(0-x)\bigg]\ \text{Tr}\bigg[\ga S_u^{ba^\prime}(y-x)\gb S_u^{\prime\ ab^\prime}(y-x)\bigg]
\nonumber	\\
	&+\bigg[S_d^{ce}(y-0) \s S_d^{ec^\prime}(0-x)\bigg]
	\text{Tr}\bigg[\ga S_u^{bb^\prime}(y-x) \gb S_u^{\prime\ aa^\prime}(y-x)\bigg],
	\label{isoscalar}
\end{align}
where $S' = C S^T C$ represent the charge conjugated quark propagator.

\vspace{0.3cm}
\underline	{\textbf{\textit{{ Isovector case:}}}}
	\allowdisplaybreaks
		\begin{align}
\Pi^{\text{isovector}}_{\alpha\mu\nu\beta}(x,y)&=
	4\ \bigg[S_u^{c c^\prime}(y-x)\bigg] \text{Tr}\bigg[\ga S_u^{ae}(y-0) \s S_u^{e a^\prime}(0-x) \gb S_d^{\prime\ b b^\prime}(y-x)\bigg]
\nonumber\\
&-4\bigg[S_u^{c a^\prime}(y-x)\gb S_d^{\prime\ b b^\prime}(y-x)\ga S_u^{ae}(y-0) \s S_u^{e c^\prime}(0-x)\bigg]
\nonumber\\
&-4\bigg[S_u^{ce}(y-0)\s S_u^{e a^\prime}(0-x)\gb S_d^{\prime\ b b^\prime}(y-x)\ga S_u^{a c^\prime}(y-x)\bigg]
\nonumber\\
&+4\bigg[S_u^{ce}(y-0)\s S_u^{ec^\prime}(0-x)\bigg]\ \text{Tr}\bigg[\ga S_u^{aa^\prime}(y-x) \gb S_d^{\prime\ b b^\prime}(y-x)\bigg]
\nonumber\\
&-2\bigg[ S_u^{cb^\prime}(y-x)\gb S_u^{\prime\ ea^\prime}(0-x)\s S_u^{\prime\ ae}(y-0)\ga S_d^{bc^\prime}(y-x)\bigg]
\nonumber\\
&+2\bigg[ S_u^{ca^\prime}(y-x)\gb S_u^{\prime\ eb^\prime}(0-x)\s S_u^{\prime\ ae}(y-0)\ga S_d^{bc^\prime}(y-x)\bigg]
\nonumber\\
&+2\bigg[S_u^{ce}(y-0) \s S_u^{ea^\prime}(0-x) \gb S_u^{\prime\ ab^\prime}(y-x)\ga S_d^{bc^\prime}(y-x)\bigg]
\nonumber\\
&-2\bigg[S_u^{ce}(y-0) \s S_u^{eb^\prime}(0-x) \gb S_u^{\prime\ aa^\prime}(y-x)\ga S_d^{bc^\prime}(y-x)\bigg]
\nonumber\\
&+4\bigg[S_u^{ca^\prime}(y-x) \gb S_d^{\prime\ eb^\prime}(0-x)\s S_d^{\prime\ be}(y-0)\ga S_u^{ac^\prime}(y-x)\bigg]
\nonumber\\
&-4\bigg[S_u^{cc^\prime}(y-x)\bigg]\ \text{Tr}\bigg[\ga S_d^{be}(y-0) \s S_d^{eb^\prime}(0-x) \gb S_u^{\prime\ aa^\prime}(y-x)\bigg]
\nonumber\\
&-2\bigg[ S_u^{ca^\prime}(y-x)\gb S_u^{\prime\ ab^\prime}(y-x)\ga S_d^{be}(y-0) \s S_d^{ec^\prime}(0-x)\bigg]
\nonumber\\
&+2\bigg[S_u^{cb^\prime}(y-x)\gb S_u^{\prime\ aa^\prime}(y-x)\ga S_d^{be}(y-0)\s S_d^{ec^\prime}(0-x)\bigg]
\nonumber\\
&-2\bigg[S_d^{cb^\prime}(y-x)\gb S_u^{\prime\ ea^\prime}(0-x)\s S_u^{\prime\ ae}(y-0)\ga S_u^{bc^\prime}(y-x)\bigg]
\nonumber\\
&+2\bigg[S_d^{cb^\prime}(y-x)\gb S_u^{\prime\ ba^\prime}(y-x)\ga S_u^{ ae}(y-0)\s S_u^{ec^\prime}(0-x)\bigg]
\nonumber\\
&+2\bigg[S_d^{cb^\prime}(y-x)\gb S_u^{\prime\ ea^\prime}(0-x)\s S_u^{\prime\ be}(y-0)\ga S_u^{ac^\prime}(y-x)\bigg]
\nonumber\\
&-2\bigg[S_d^{cb^\prime}(y-x)\gb S_u^{\prime\ aa^\prime}(y-x)\ga S_u^{ be}(y-0)\s S_u^{ec^\prime}(0-x)\bigg]
\nonumber\\
&+\bigg[S_d^{cc^\prime}(y-x)\bigg]\ \text{Tr}\bigg[\ga S_u^{bb^\prime}(y-x)\gb S_u^{\prime\ ea^\prime}(0-x)\s S_u^{\prime\ ae}(y-0)\bigg]
\nonumber\\
&-\bigg[S_d^{cc^\prime}(y-x)\bigg]\ \text{Tr}\bigg[\ga S_u^{ba^\prime}(y-x)\gb S_u^{\prime\ eb^\prime}(0-x)\s S_u^{\prime\ ae}(y-0)\bigg]
\nonumber\\
&-\bigg[S_d^{cc^\prime}(y-x)\bigg]\ \text{Tr}\bigg[\ga S_u^{be}(y-0)\s S_u^{ ea^\prime}(0-x)\gb S_u^{\prime\ ab^\prime}(y-x)\bigg]
\nonumber\\
&+\bigg[S_d^{cc^\prime}(y-x)\bigg]\ \text{Tr}\bigg[\ga S_u^{be}(y-0)\s S_u^{ eb^\prime}(0-x)\gb S_u^{\prime\ aa^\prime}(y-x)\bigg]
\nonumber\\
&-2\bigg[S_d^{ce}(y-0)\s S_d^{eb^\prime}(0-x)\gb S_u^{\prime\ ba^\prime}(y-x)\ga S_u^{ac^\prime}(y-x)\bigg]
\nonumber\\
&+2\bigg[S_d^{ce}(y-0)\s S_d^{eb^\prime}(0-x)\gb S_u^{\prime\ aa^\prime}(y-x)\ga S_u^{bc^\prime}(y-x)\bigg]
\nonumber\\
&+\bigg[S_d^{ce}(y-0)\s S_d^{ec^\prime}(0-x)\bigg]\text{Tr}\bigg[\ga S_u^{ba^\prime}(y-x)\gb S_u^{\prime\ ab^\prime}(y-x)\bigg]
\nonumber\\
&-\bigg[S_d^{ce}(y-0) \s S_d^{ec^\prime}(0-x)\bigg]
			\text{Tr}\bigg[\ga S_u^{bb^\prime}(y-x) \gb S_u^{\prime\ aa^\prime}(y-x)\bigg].
			\label{isovector}
\end{align}
The light-quark propagator $S^{ab}_\psi(x)$ appearing in Eqs. \eqref{isoscalar} and \eqref{isovector} is defined as \cite{Agaev:2020zad}
\begin{equation}
	\begin{split}
		S_\psi^{ab}(x) &= i \delta^{ab} \frac{\slashed{x}}{2\pi^2 x^4} 
		- \delta^{ab} \frac{m_\psi}{4 \pi^2 x^2} 
		-\delta^{ab} \frac{\langle \bar{\psi} \psi \rangle}{12}\bigg(1-\frac{m_\psi \slashed{x}}{4}\bigg)
		- \delta^{ab} \frac{x^2}{192}\left\langle  \bar{\psi} g_s \sigma G \psi \right\rangle\bigg(1-i\frac{m_\psi \slashed{x}}{6}\bigg)
		\\
		& \quad 
		- i \frac{g_s G^{ab}_{\alpha \beta}}{32 \pi^2 x^2} \left[ \slashed{x} \sigma^{\alpha \beta} + \sigma^{\alpha \beta} \slashed{x} \right]
		- i \delta^{ab} \frac{x^2 \slashed{x} g_s^2 \langle \bar{\psi} \psi \rangle^2}{7776} 
	 + \cdots,
		\label{propagator}
	\end{split}
\end{equation}
where $m_\psi$ denotes the mass of the light quark.
$\langle \bar{\psi} \psi \rangle$, $\langle G^2 \rangle$ and $\langle \bar{\psi} g_s \sigma G \psi\rangle$ represent the two-quark, two-gluon, and mixed quark-gluon condensates, respectively. $g_s$ is the strong coupling constant and $\sigma^{\alpha\beta}$  denotes the anti-symmetric combination of gamma matrices.
The gluon strength field tensor $G^{ab}_{\alpha\beta}$ is define as
\begin{equation}
	G_{\alpha\beta}^{ab} = G_{\alpha\beta}^{A} T_{A}^{ab}, \qquad 
	T_{A} = \frac{1}{2} \lambda_{A}, \qquad
	G^2 = G^{A}_{\alpha\beta} G^{A}_{\alpha\beta},
\end{equation}
where $a, b = 1, 2, 3$ and $A = 1,\dots, 8$ denote the color degrees of freedom of quark and gluon fields, respectively, and $T_A$ represent the SU(3) generators in the fundamental representation, written as half of the Gell-Mann matrices, $\lambda^A/2$.		
The vacuum expectation value of the gluon fields is given by
\begin{equation}
	\langle G_{\alpha^\prime\beta^\prime}^{A_1} G_{\alpha\beta}^{A_2} \rangle
	= \frac{\delta^{A_1 A_2}}{96} \, \langle G^2 \rangle 
	\left( g_{\alpha^\prime\alpha} g_{\beta^\prime\beta} - g_{\alpha^\prime\beta} g_{\beta^\prime\alpha} \right),
	\label{gluon}
\end{equation}
while the color contraction satisfies the SU($3$) identity
\begin{equation}
	T_A^{ab} \, T_A^{cd} = \frac{1}{2} \left( \delta^{ad} \delta^{bc} - \frac{1}{3} \delta^{ab} \delta^{cd} \right).
	\label{generator}
\end{equation}
By applying Eqs.~\eqref{gluon} and \eqref{generator}, the gluon condensate with explicit color and Lorentz indices can be written as
\begin{equation}
\langle0|G^{ab}_{\alpha'\beta'}(0) G^{cd}_{\alpha\beta}(0)|0\rangle=\langle0| G_{\alpha^\prime\beta^\prime}^{A_1}(0) \, T_{A_1}^{ab} \, G_{\alpha\beta}^{A_2}(0) \, T_{A_2}^{cd} |0\rangle
	= \frac{\langle G^2 \rangle}{192} 
	\left( g_{\alpha^\prime\alpha} g_{\beta^\prime\beta} - g_{\alpha^\prime\beta} g_{\beta^\prime\alpha} \right)
\left( \delta^{ad} \delta^{bc} - \frac{1}{3} \delta^{ab} \delta^{cd} \right).
\end{equation}	
\subsection{Borel transformation}
In this subsection, we list the Borel transformations applied with respect to the $p^2$, with similar expressions applied for $p^{\prime 2}$.
	\allowdisplaybreaks
\begin{align}
\begin{array}{l@{\hspace{0.01cm}}l}
	\mathcal{B}_{M_1^2} e^{-\frac{p^2}{4\alpha}}= \delta\Big(\frac{1}{M_1^2}-\frac{1}{4\alpha}\Big),
\\
\\
	\mathcal{B}_{M_1^2}\left( p^2 e^{-\frac{p^2}{4\alpha}}\right) =4\alpha^2 \frac{d}{d\alpha}\delta\Big(\frac{1}{M_1^2}-\frac{1}{4\alpha}\Big), 
	\\
	\\
	\mathcal{B}_{M_1^2}\left( p^4 e^{-\frac{p^2}{4\alpha}}\right) =16\alpha^4 \frac{d^2}{d\alpha^2}\delta\Big(\frac{1}{M_1^2}-\frac{1}{4\alpha}\Big)+32\alpha^3 \frac{d}{d\alpha}\delta\Big(\frac{1}{M_1^2}-\frac{1}{4\alpha}\Big), 
	\\
	\\
	\mathcal{B}_{M_1^2}\left( p^6 e^{-\frac{p^2}{4\alpha}}\right) =64\alpha^6 \frac{d^3}{d\alpha^3}\delta\Big(\frac{1}{M_1^2}-\frac{1}{4\alpha}\Big)+384\alpha^5 \frac{d^2}{d\alpha^2}\delta\Big(\frac{1}{M_1^2}-\frac{1}{4\alpha}\Big)+384\alpha^4 \frac{d}{d\alpha}\delta\Big(\frac{1}{M_1^2}-\frac{1}{4\alpha}\Big).\\
\end{array}
\label{borel}
\end{align}

\begin{acknowledgments} 
	Z. Asmaee and K. Azizi are thankful to the Iran National Science Foundation (INSF) for the financial support
	provided for this research under the grant number 4036353.

\end{acknowledgments}


\begin{thebibliography}{999}
	%\cite{Aliev:2008cs}
	\bibitem{Aliev:2008cs}
	T.~M.~Aliev, K.~Azizi, A.~Ozpineci and M.~Savci,
 Nucleon Electromagnetic Form Factors in QCD,
 \href{https://doi.org//10.1103/PhysRevD.77.114014}{Phys. Rev. D \textbf{77}, 114014 (2008)},
 \href{https://arxiv.org/abs/0802.3008}{\color{teal}{[arXiv:0802.3008 [hep-ph]]}}.
	%26 citations counted in INSPIRE as of 08 Nov 2025 
	
	%\cite{Er:2022cxx}
	\bibitem{Er:2022cxx}
	N.~Er and K.~Azizi,
	Spectroscopic parameters and electromagnetic form factor of kaon in vacuum and a dense medium,
	\href{https://doi.org//10.1140/epjc/s10052-022-10333-w}{Eur. Phys. J. C \textbf{82}, 397 (2022)},
	\href{https://arxiv.org/abs/2202.01504}{\color{teal}{	[arXiv:2202.01504 [hep-ph]]}}.
	%8 citations counted in INSPIRE as of 08 Nov 2025
	
	%\cite{Sundu:2018uyi}
	\bibitem{Sundu:2018uyi}
	H.~Sundu, B.~Barsbay, S.~S.~Agaev and K.~Azizi,
	Probing an axial-vector tetraquark Z$_{s}$ via its semileptonic decay Z$_{s}$ $ \rightarrow$ X(4274) $ \overline{{l}}$ $ \nu_{l}^{}$,
	\href{https://doi.org//10.1140/epja/i2018-12552-0}{Eur. Phys. J. A \textbf{54}, 124 (2018)},
	\href{https://arxiv.org/abs/1804.04525}{\color{teal}{[arXiv:1804.04525 [hep-ph]]}}.
	%9 citations counted in INSPIRE as of 08 Nov 2025
	
%\cite{Sun:2020wfo}
\bibitem{Sun:2020wfo}
B.~D.~Sun and Y.~B.~Dong,
Gravitational form factors of $\rho$ meson with a light-cone constituent quark model,
\href{https://doi.org//10.1103/PhysRevD.101.096008}{Phys. Rev. D \textbf{101},  096008 (2020)},
\href{https://arxiv.org/abs/2002.02648}{\color{teal}{[arXiv:2002.02648 [hep-ph]]}}.
%31 citations counted in INSPIRE as of 26 Oct 2025

%\cite{Fu:2023dea}
\bibitem{Fu:2023dea}
D.~Fu, B.~D.~Sun and Y.~Dong,
Generalized parton distributions of {\ensuremath{\Delta}} resonance in a diquark spectator approach,
\href{https://doi.org//10.1103/PhysRevD.107.116021}{Phys. Rev. D \textbf{107}, 116021 (2023)},
\href{https://arxiv.org/abs/2305.02680}{\color{teal}{[arXiv:2305.02680 [hep-ph]]}}.
%8 citations counted in INSPIRE as of 26 Oct 2025

%\cite{Kumano:2017lhr}
\bibitem{Kumano:2017lhr}
S.~Kumano, Q.~T.~Song and O.~V.~Teryaev,
Hadron tomography by generalized distribution amplitudes in pion-pair production process $\gamma^* \gamma \rightarrow \pi^0 \pi^0 $ and gravitational form factors for pion,
\href{https://doi.org//10.1103/PhysRevD.97.014020}{Phys. Rev. D \textbf{97}, 014020 (2018)},
\href{https://arxiv.org/abs/1711.08088}{\color{teal}{[arXiv:1711.08088 [hep-ph]]}}.
%154 citations counted in INSPIRE as of 26 Oct 2025

%\cite{Bincer:1959tz}
\bibitem{Bincer:1959tz}
A.~M.~Bincer,
Electromagnetic structure of the nucleon,
\href{https://doi.org//10.1103/PhysRev.118.855}{Phys. Rev. \textbf{118}, 855-863 (1960)}.
%162 citations counted in INSPIRE as of 26 Oct 2025

%\cite{Keiner:1996at}
\bibitem{Keiner:1996at}
V.~Keiner,
A Covariant diquark - quark model of the nucleon in the Salpeter approach,
\href{https://doi.org//10.1103/PhysRevC.54.3232}{Phys. Rev. C \textbf{54}, 3232-3239 (1996)},
\href{https://arxiv.org/abs/hep-ph/9603226}{\color{teal}{[arXiv:hep-ph/9603226 [hep-ph]]}}.
%25 citations counted in INSPIRE as of 26 Oct 2025
%9
%\cite{Kim:2012ts}
\bibitem{Kim:2012ts}
H.~C.~Kim, P.~Schweitzer and U.~Yakhshiev,
Energy-momentum tensor form factors of the nucleon in nuclear matter,
\href{https://doi.org//10.1016/j.physletb.2012.10.055}{Phys. Lett. B \textbf{718}, 625-631 (2012)},
\href{https://arxiv.org/pdf/1205.5228v2}{\color{teal}{[arXiv:1205.5228v2 [hep-ph]]}}.
%78 citations counted in INSPIRE as of 26 Oct 2025

%\cite{Cosyn:2018thq}
\bibitem{Cosyn:2018thq}
W.~Cosyn, A.~Freese and B.~Pire,
Polynomiality sum rules for generalized parton distributions of spin-1 targets,
\href{https://doi.org//10.1103/PhysRevD.99.094035}{Phys. Rev. D \textbf{99}, 094035 (2019)},
\href{https://arxiv.org/abs/1812.01511}{\color{teal}{[arXiv:1812.01511 [hep-ph]]}}.
%15 citations counted in INSPIRE as of 26 Oct 2025

%\cite{Sun:2017gtz}
\bibitem{Sun:2017gtz}
B.~D.~Sun and Y.~B.~Dong,
$\rho$ meson unpolarized generalized parton distributions with a light-front constituent quark model,
\href{https://doi.org//10.1103/PhysRevD.96.036019}{Phys. Rev. D \textbf{96},  036019 (2017)},
\href{https://arxiv.org/abs/1707.03972}{\color{teal}{[arXiv:1707.03972 [hep-ph]]}}.
%39 citations counted in INSPIRE as of 26 Oct 2025

%\cite{Dong:2013rk}
\bibitem{Dong:2013rk}
Y.~Dong and C.~Liang,
Generalized parton distribution functions of a deuteron in a phenomenological Lagrangian approach,
\href{https://doi.org//10.1088/0954-3899/40/2/025001}{J. Phys. G \textbf{40}, 025001 (2013)}.
%12 citations counted in INSPIRE as of 26 Oct 2025
%\cite{Polyakov:2019lbq}
\bibitem{Polyakov:2019lbq}
M.~V.~Polyakov and B.~D.~Sun,
Gravitational form factors of a spin one particle,
\href{https://doi.org//10.1103/PhysRevD.100.036003}{Phys. Rev. D \textbf{100}, 036003 (2019)},
\href{https://arxiv.org/abs/1903.02738}{\color{teal}{[arXiv:1903.02738 [hep-ph]]}}.
%58 citations counted in INSPIRE as of 26 Oct 2025

%\cite{Alexandrou:2008bn}
\bibitem{Alexandrou:2008bn}
C.~Alexandrou, T.~Korzec, G.~Koutsou, T.~Leontiou, C.~Lorce, J.~W.~Negele, V.~Pascalutsa, A.~Tsapalis and M.~Vanderhaeghen,
Delta-baryon electromagnetic form factors in lattice QCD,
\href{https://doi.org//10.1103/PhysRevD.79.014507}{Phys. Rev. D \textbf{79}, 014507 (2009)},
\href{https://arxiv.org/abs/0810.3976}{\color{teal}{[arXiv:0810.3976 [hep-lat]]}}.
%81 citations counted in INSPIRE as of 26 Oct 2025

%\cite{Alexandrou:2010jv}
\bibitem{Alexandrou:2010jv}
C.~Alexandrou, T.~Korzec, G.~Koutsou, J.~W.~Negele and Y.~Proestos,
The Electromagnetic form factors of the $\Omega^-$ in lattice QCD,
\href{https://doi.org//10.1103/PhysRevD.82.034504}{Phys. Rev. D \textbf{82}, 034504 (2010)},
\href{https://arxiv.org/abs/1006.0558}{\color{teal}{[arXiv:1006.0558 [hep-lat]]}}.
%41 citations counted in INSPIRE as of 26 Oct 2025

%\cite{Boinepalli:2009sq}
\bibitem{Boinepalli:2009sq}
S.~Boinepalli, D.~B.~Leinweber, P.~J.~Moran, A.~G.~Williams, J.~M.~Zanotti and J.~B.~Zhang,
Precision electromagnetic structure of decuplet baryons in the chiral regime,
\href{https://doi.org//10.1103/PhysRevD.80.054505}{Phys. Rev. D \textbf{80}, 054505 (2009)},
\href{https://arxiv.org/abs/0902.4046}{\color{teal}{[arXiv:0902.4046 [hep-lat]]}}.
%65 citations counted in INSPIRE as of 26 Oct 2025

%\cite{lattice}
\bibitem{lattice}
C.~Aubin, K.~Orginos, V.~Pascalutsa, and M.~Vanderhaeghen, Lattice calculation of the magnetic moments of   and    baryons with dynamical clover fermions, \href{https://doi.org/10.1103/PhysRevD.79.051502}{Phys. Rev. D \textbf{79}, 051502 (2009)},
\href{https://doi.org/10.48550/arXiv.0811.2440}{\color{teal}{[arXiv:0811.2440 [hep-lat]]}}.

%\cite{Leinweber:1992hy}
\bibitem{Leinweber:1992hy}
D.~B.~Leinweber, T.~Draper and R.~M.~Woloshyn,
Decuplet baryon structure from lattice QCD,
\href{https://doi.org//10.1103/PhysRevD.46.3067}{Phys. Rev. D \textbf{46}, 3067-3085 (1992)},
\href{https://arxiv.org/abs/hep-lat/9208025 }{\color{teal}{[arXiv:hep-lat/9208025 [hep-lat]]}}.
%192 citations counted in INSPIRE as of 26 Oct 2025

%\cite{Lee:2005ds}
\bibitem{Lee:2005ds}
F.~X.~Lee, R.~Kelly, L.~Zhou and W.~Wilcox,
Baryon magnetic moments in the background field method,
\href{https://doi.org//10.1016/j.physletb.2005.08.106}{Phys. Lett. B \textbf{627}, 71-76 (2005)},
\href{https://arxiv.org/abs/hep-lat/0509067}{\color{teal}{[arXiv:hep-lat/0509067 [hep-lat]]}}.
%93 citations counted in INSPIRE as of 26 Oct 2025

%\cite{Oh:1995hn}
\bibitem{Oh:1995hn}
Y.~s.~Oh,
Electric quadrupole moments of the decuplet baryons in the Skyrme model,
\href{https://doi.org//10.1142/S0217732395001137}{Mod. Phys. Lett. A \textbf{10}, 1027-1034 (1995)},
\href{https://arxiv.org/abs/hep-ph/9506308 }{\color{teal}{[arXiv:hep-ph/9506308 [hep-ph]]}}.
%25 citations counted in INSPIRE as of 27 Oct 2025

%\cite{Geng:2009ys}
\bibitem{Geng:2009ys}
L.~S.~Geng, J.~Martin Camalich and M.~J.~Vicente Vacas,
Electromagnetic structure of the lowest-lying decuplet resonances in covariant chiral perturbation theory,
\href{https://doi.org//10.1103/PhysRevD.80.034027}{Phys. Rev. D \textbf{80}, 034027 (2009)},
\href{https://arxiv.org/abs/0907.0631}{\color{teal}{[arXiv:0907.0631 [hep-ph]]}}.
%80 citations counted in INSPIRE as of 27 Oct 2025

%\cite{Li:2016ezv}
\bibitem{Li:2016ezv}
H.~S.~Li, Z.~W.~Liu, X.~L.~Chen, W.~Z.~Deng and S.~L.~Zhu,
Magnetic moments and electromagnetic form factors of the decuplet baryons in chiral perturbation theory,
\href{https://doi.org//10.1103/PhysRevD.95.076001}{Phys. Rev. D \textbf{95}, 076001 (2017)},
\href{https://arxiv.org/abs/1608.04617 }{\color{teal}{[arXiv:1608.04617 [hep-ph]]}}.
%19 citations counted in INSPIRE as of 27 Oct 2025

%\cite{Krivoruchenko:1991pm}
\bibitem{Krivoruchenko:1991pm}
M.~I.~Krivoruchenko and M.~M.~Giannini,
Quadrupole moments of the decuplet baryons,
\href{https://doi.org//10.1103/PhysRevD.43.3763}{Phys. Rev. D \textbf{43}, 3763-3765 (1991)}.
%43 citations counted in INSPIRE as of 27 Oct 2025

%\cite{Berger:2004yi}
\bibitem{Berger:2004yi}
K.~Berger, R.~F.~Wagenbrunn and W.~Plessas,
Covariant baryon charge radii and magnetic moments in a chiral constituent quark model,
\href{https://doi.org//10.1103/PhysRevD.70.094027}{Phys. Rev. D \textbf{70}, 094027 (2004)},
\href{https://arxiv.org/abs/nucl-th/0407009}{\color{teal}{[arXiv:nucl-th/0407009 [nucl-th]]}}.
%49 citations counted in INSPIRE as of 27 Oct 2025

%\cite{Schlumpf:1993rm}
\bibitem{Schlumpf:1993rm}
F.~Schlumpf,
Magnetic moments of the baryon decuplet in a relativistic quark model,
\href{https://doi.org//10.1103/PhysRevD.48.4478}{Phys. Rev. D \textbf{48}, 4478-4480 (1993)},
\href{https://arxiv.org/abs/hep-ph/9305293 }{\color{teal}{[arXiv:hep-ph/9305293 [hep-ph]]}}.
%104 citations counted in INSPIRE as of 27 Oct 2025

%\cite{Aliev:2009jt}
\bibitem{Aliev:2009jt}
T.~M.~Aliev, K.~Azizi and A.~Ozpineci,
Radiative Decays of the Heavy Flavored Baryons in Light Cone QCD Sum Rules,
\href{https://doi.org//10.1103/PhysRevD.79.056005}{Phys. Rev. D \textbf{79}, 056005 (2009)},
\href{https://arxiv.org/abs/0901.0076}{\color{teal}{[arXiv:0901.0076 [hep-ph]]}}.

%\cite{Aliev:2010uy}
\bibitem{Aliev:2010uy}
T.~M.~Aliev, K.~Azizi and M.~Savci,
Analysis of the $\Lambda_{b}\rightarrow \Lambda \ell^+\ell^- $ decay in QCD,
\href{https://doi.org//10.1103/PhysRevD.81.056006}{Phys. Rev. D \textbf{81}, 056006 (2010)},
\href{https://arxiv.org/abs/1001.0227}{\color{teal}{[arXiv:1001.0227 [hep-ph]]}}.

%\cite{Aliev:2009pd}
\bibitem{Aliev:2009pd}
T.~M.~Aliev, K.~Azizi and M.~Savci,
Electric Quadrupole and Magnetic Octupole Moments of the Light Decuplet Baryons Within Light Cone QCD Sum Rules,
\href{https://doi.org//10.1016/j.physletb.2009.10.026}{Phys. Lett. B \textbf{681}, 240-246 (2009)},
\href{https://arxiv.org/abs/0904.2485}{\color{teal}{[arXiv:0904.2485 [hep-ph]]}}.
%28 citations counted in INSPIRE as of 27 Oct 2025

%\cite{Lee:1997jk}
\bibitem{Lee:1997jk}
F.~X.~Lee,
Determination of decuplet baryon magnetic moments from QCD sum rules,
\href{https://doi.org//10.1103/PhysRevD.57.1801}{Phys. Rev. D \textbf{57}, 1801-1821 (1998)},
\href{https://arxiv.org/abs/hep-ph/9708323}{\color{teal}{[arXiv:hep-ph/9708323 [hep-ph]]}}.
%69 citations counted in INSPIRE as of 27 Oct 2025

%\cite{Azizi:2009egn}
\bibitem{Azizi:2009egn}
K.~Azizi,
Magnetic Dipole, Electric Quadrupole and Magnetic Octupole Moments of the Delta Baryons in Light Cone QCD Sum Rules,
\href{https://doi.org//10.1140/epjc/s10052-009-0988-0}{Eur. Phys. J. C \textbf{61}, 311-319 (2009)},
\href{https://arxiv.org/abs/0811.2670}{\color{teal}{[arXiv:0811.2670 [hep-ph]]}}.
%31 citations counted in INSPIRE as of 27 Oct 2025


%\cite{Wagner:2000ii}
\bibitem{Wagner:2000ii}
G.~Wagner, A.~J.~Buchmann and A.~Faessler,
Electromagnetic properties of decuplet hyperons in a chiral quark model with exchange currents,
\href{https://doi.org//10.1088/0954-3899/26/3/306}{J. Phys. G \textbf{26}, 267-293 (2000)}.
%38 citations counted in INSPIRE as of 27 Oct 2025

%\cite{Kim:2019gka}
\bibitem{Kim:2019gka}
J.~Y.~Kim and H.~C.~Kim,
Electromagnetic form factors of the baryon decuplet with flavor SU(3) symmetry breaking,
\href{https://doi.org//10.1140/epjc/s10052-019-7079-7}{Eur. Phys. J. C \textbf{79}, 570 (2019)},
\href{https://arxiv.org/abs/1905.04017}{\color{teal}{[arXiv:1905.04017 [hep-ph]]}}.
%19 citations counted in INSPIRE as of 27 Oct 2025

%\cite{Kim:2020lrs}
\bibitem{Kim:2020lrs}
J.~Y.~Kim and B.~D.~Sun,
Gravitational form factors of a baryon with spin-3/2,
\href{https://doi.org//10.1140/epjc/s10052-021-08852-z}{Eur. Phys. J. C \textbf{81}, 85 (2021)},
\href{https://arxiv.org/abs/2011.00292}{\color{teal}{[arXiv:2011.00292 [hep-ph]]}}.
%41 citations counted in INSPIRE as of 27 Oct 2025

%\cite{Wang:2023bjp}
\bibitem{Wang:2023bjp}
J.~Wang, D.~Fu and Y.~Dong,
Form factors of decuplet baryons in a covariant quark{\textendash}diquark approach,
\href{https://doi.org//10.1140/epjc/s10052-024-12406-4}{Eur. Phys. J. C \textbf{84}, 79 (2024)},
\href{https://arxiv.org/abs/2311.07149 }{\color{teal}{[arXiv:2311.07149 [hep-ph]]}}.
%8 citations counted in INSPIRE as of 27 Oct 2025

%\cite{Dehghan:2023ytx}
\bibitem{Dehghan:2023ytx}
Z.~Dehghan, K.~Azizi and U.~{\"O}zdem,
Gravitational form factors of the {\ensuremath{\Delta}} baryon via QCD sum rules,
\href{https://doi.org//10.1103/PhysRevD.108.094037}{Phys. Rev. D \textbf{108}, 094037 (2023)},
\href{https://arxiv.org/abs/2307.14880}{\color{teal}{[arXiv:2307.14880 [hep-ph]]}}.
%12 citations counted in INSPIRE as of 27 Oct 2025

%\cite{Dehghan:2025ncw}
\bibitem{Dehghan:2025ncw}
Z.~Dehghan, F.~Almaksusi and K.~Azizi,
Mechanical properties of proton using flavor-decomposed gravitational form factors,
\href{https://doi.org//10.1007/JHEP06(2025)025}{JHEP \textbf{06}, 025 (2025)},
\href{https://arxiv.org/abs/2502.16689}{\color{teal}{[arXiv:2502.16689 [hep-ph]]}}.
%15 citations counted in INSPIRE as of 27 Oct 2025


%\cite{Dehghan:2025eov}
\bibitem{Dehghan:2025eov}
Z.~Dehghan and K.~Azizi,
Mechanical properties of the {\ensuremath{\Omega}}- baryon from gravitational form factors,
\href{https://doi.org//10.1103/x6f5-cbcc}{Phys. Rev. D \textbf{112}, 054014 (2025)},
\href{https://arxiv.org/abs/2507.14840}{\color{teal}{[arXiv:2507.14840 [hep-ph]]}}.
%3 citations counted in INSPIRE as of 27 Oct 2025

%\cite{kucukarslan:2016xhx}
\bibitem{kucukarslan:2016xhx}
A.~kucukarslan, U.~Ozdem and A.~Ozpineci,
Tensor form factors of the octet hyperons in QCD,
\href{https://doi.org//10.1103/PhysRevD.94.094010}{Phys. Rev. D \textbf{94}, 094010 (2016)},
\href{https://arxiv.org/abs/1610.08358}{\color{teal}{[arXiv:1610.08358 [hep-ph]]}}.
%6 citations counted in INSPIRE as of 27 Oct 2025

%\cite{Aliev:2011ku}
\bibitem{Aliev:2011ku}
T.~M.~Aliev, K.~Azizi and M.~Savci,
Nucleon tensor form factors induced by isovector and isoscalar currents in QCD,
\href{https://doi.org//10.1103/PhysRevD.84.076005}{Phys. Rev. D \textbf{84}, 076005 (2011)},
\href{https://arxiv.org/abs/1108.2019}{\color{teal}{[arXiv:1108.2019 [hep-ph]]}}.
%16 citations counted in INSPIRE as of 27 Oct 2025

%\cite{Gutsche:2016xff}
\bibitem{Gutsche:2016xff}
T.~Gutsche, M.~A.~Ivanov, J.~G.~Korner, S.~Kovalenko and V.~E.~Lyubovitskij,
Nucleon tensor form factors in a relativistic confined quark model,
\href{https://doi.org//10.1103/PhysRevD.94.114030}{Phys. Rev. D \textbf{94}, 114030 (2016)},
\href{https://arxiv.org/abs/1608.00420}{\color{teal}{[arXiv:1608.00420 [hep-ph]]}}.
%7 citations counted in INSPIRE as of 28 Oct 2025

%\cite{Azizi:2019ytx}
\bibitem{Azizi:2019ytx}
K.~Azizi and U.~{\"O}zdem,
Nucleon{\textquoteright}s energy{\textendash}momentum tensor form factors in light-cone QCD,
\href{https://doi.org//10.1140/epjc/s10052-020-7676-5}{Eur. Phys. J. C \textbf{80}, 104 (2020)},
\href{https://arxiv.org/abs/1908.06143}{\color{teal}{[arXiv:1908.06143 [hep-ph]]}}.
%50 citations counted in INSPIRE as of 28 Oct 2025

%\cite{Ozdem:2020vpt}
\bibitem{Ozdem:2020vpt}
U.~{\"O}zdem,
Isovector and isoscalar tensor form factors of $N(1535) \rightarrow N$ transition in light-cone QCD,
\href{https://doi.org//10.1103/PhysRevD.102.014001}{Phys. Rev. D \textbf{102}, 014001 (2020)},
\href{https://arxiv.org/abs/2004.12312}{\color{teal}{[arXiv:2004.12312 [hep-ph]]}}.
%1 citations counted in INSPIRE as of 28 Oct 2025

%\cite{Ozdem:2021zbn}
\bibitem{Ozdem:2021zbn}
U.~{\"O}zdem,
Tensor form factors of N(1535) state via light-cone QCD,
\href{https://doi.org//10.1016/j.cjph.2021.04.018}{Chin. J. Phys. \textbf{72}, 93-99 (2021)}.
%0 citations counted in INSPIRE as of 28 Oct 2025

%\cite{Ralston:1979ys}
\bibitem{Ralston:1979ys}
J.~P.~Ralston and D.~E.~Soper,
Production of Dimuons from High-Energy Polarized Proton Proton Collisions,
\href{https://doi.org//10.1016/0550-3213(79)90082-8}{Nucl. Phys. B \textbf{152}, 109 (1979)}.
%769 citations counted in INSPIRE as of 28 Oct 2025

%\cite{Jaffe:1991kp}
\bibitem{Jaffe:1991kp}
R.~L.~Jaffe and X.~D.~Ji,
Chiral odd parton distributions and polarized Drell-Yan,
\href{https://doi.org//10.1103/PhysRevLett.67.552}{Phys. Rev. Lett. \textbf{67}, 552-555 (1991)}.
%631 citations counted in INSPIRE as of 28 Oct 2025

%\cite{Jaffe:1991ra}
\bibitem{Jaffe:1991ra}
R.~L.~Jaffe and X.~D.~Ji,
Chiral odd parton distributions and Drell-Yan processes,
\href{https://doi.org//10.1016/0550-3213(92)90110-W}{Nucl. Phys. B \textbf{375}, 527-560 (1992)}.
%706 citations counted in INSPIRE as of 28 Oct 2025

%\cite{Barone:2001sp}
\bibitem{Barone:2001sp}
V.~Barone, A.~Drago and P.~G.~Ratcliffe,
Transverse polarisation of quarks in hadrons,
\href{https://doi.org//10.1016/S0370-1573(01)00051-5}{Phys. Rept. \textbf{359}, 1-168 (2002)},
\href{https://arxiv.org/abs/hep-ph/0104283}{\color{teal}{[arXiv:hep-ph/0104283 [hep-ph]]}}.
%632 citations counted in INSPIRE as of 28 Oct 2025

%\cite{Anselmino:2007fs}
\bibitem{Anselmino:2007fs}
M.~Anselmino, M.~Boglione, U.~D'Alesio, A.~Kotzinian, F.~Murgia, A.~Prokudin and C.~Turk,
Transversity and Collins functions from SIDIS and e+ e- data,
\href{https://doi.org//10.1103/PhysRevD.75.054032}{Phys. Rev. D \textbf{75}, 054032 (2007)},
\href{https://arxiv.org/abs/hep-ph/0701006}{\color{teal}{[arXiv:hep-ph/0701006 [hep-ph]]}}.
%467 citations counted in INSPIRE as of 28 Oct 2025

%\cite{Anselmino:2008jk}
\bibitem{Anselmino:2008jk}
M.~Anselmino, M.~Boglione, U.~D'Alesio, A.~Kotzinian, F.~Murgia, A.~Prokudin and S.~Melis,
Update on transversity and Collins functions from SIDIS and e+ e- data,
\href{https://doi.org//10.1016/j.nuclphysbps.2009.03.117}{Nucl. Phys. B Proc. Suppl. \textbf{191}, 98-107 (2009)},
\href{https://arxiv.org/abs/0812.4366 }{\color{teal}{[arXiv:0812.4366 [hep-ph]]}}.
%290 citations counted in INSPIRE as of 28 Oct 2025

%\cite{Anselmino:2013vqa}
\bibitem{Anselmino:2013vqa}
M.~Anselmino, M.~Boglione, U.~D'Alesio, S.~Melis, F.~Murgia and A.~Prokudin,
Simultaneous extraction of transversity and Collins functions from new SIDIS and e+e- data,
\href{https://doi.org//10.1103/PhysRevD.87.094019}{Phys. Rev. D \textbf{87}, 094019 (2013)},
\href{https://arxiv.org/abs/1303.3822}{\color{teal}{[arXiv:1303.3822 [hep-ph]]}}.
%291 citations counted in INSPIRE as of 28 Oct 2025

%\cite{He:1994gz}
\bibitem{He:1994gz}
H.~x.~He and X.~D.~Ji,
The Nucleon's tensor charge,
\href{https://doi.org//10.1103/PhysRevD.52.2960}{Phys. Rev. D \textbf{52}, 2960-2963 (1995)},
\href{https://arxiv.org/abs/hep-ph/9412235}{\color{teal}{[arXiv:hep-ph/9412235 [hep-ph]]}}.
%120 citations counted in INSPIRE as of 29 Oct 2025

%\cite{Pasquini:2005dk}
\bibitem{Pasquini:2005dk}
B.~Pasquini, M.~Pincetti and S.~Boffi,
Chiral-odd generalized parton distributions in constituent quark models,
\href{https://doi.org//10.1103/PhysRevD.72.094029}{Phys. Rev. D \textbf{72}, 094029 (2005)},
\href{https://arxiv.org/abs/hep-ph/0510376}{\color{teal}{[arXiv:hep-ph/0510376 [hep-ph]]}}.
%153 citations counted in INSPIRE as of 29 Oct 2025

%\cite{Gamberg:2001qc}
\bibitem{Gamberg:2001qc}
L.~P.~Gamberg and G.~R.~Goldstein,
Flavor spin symmetry estimate of the nucleon tensor charge,
\href{https://doi.org//10.1103/PhysRevLett.87.242001}{Phys. Rev. Lett. \textbf{87}, 242001 (2001)},
\href{https://arxiv.org/abs/hep-ph/0107176}{\color{teal}{[arXiv:hep-ph/0107176 [hep-ph]]}}.
%64 citations counted in INSPIRE as of 29 Oct 2025

%\cite{He:1996wy}
\bibitem{He:1996wy}
H.~x.~He and X.~D.~Ji,
QCD sum rule calculation for the tensor charge of the nucleon,
\href{https://doi.org//10.1103/PhysRevD.54.6897}{Phys. Rev. D \textbf{54}, 6897-6902 (1996)},
\href{https://arxiv.org/abs/hep-ph/9607408}{\color{teal}{[arXiv:hep-ph/9607408 [hep-ph]]}}.
%68 citations counted in INSPIRE as of 29 Oct 2025

%\cite{Fu:2024kfx}
\bibitem{Fu:2024kfx}
D.~Fu, Y.~Dong and S.~Kumano,
Transversity generalized parton distributions in spin-3/2 particles,
\href{https://doi.org//10.1103/PhysRevD.109.096006}{Phys. Rev. D \textbf{109}, 096006 (2024)},
\href{https://arxiv.org/abs/2402.11561}{\color{teal}{[arXiv:2402.11561 [hep-ph]]}}.
%5 citations counted in INSPIRE as of 07 Dec 2025



%\cite{Diehl:2005jf}
\bibitem{Diehl:2005jf}
M.~Diehl and P.~Hagler,
Spin densities in the transverse plane and generalized transversity distributions,
\href{https://doi.org//10.1140/epjc/s2005-02342-6}{Eur. Phys. J. C \textbf{44}, 87-101 (2005)},
\href{https://arxiv.org/abs/hep-ph/0504175}{\color{teal}{[arXiv:hep-ph/0504175 [hep-ph]]}}.
%206 citations counted in INSPIRE as of 28 Oct 2025

%\cite{Burkardt:2000za}
\bibitem{Burkardt:2000za}
M.~Burkardt,
Impact parameter dependent parton distributions and off forward parton distributions for zeta ---{\ensuremath{>}} 0,
\href{https://doi.org//10.1103/PhysRevD.62.071503}{Phys. Rev. D \textbf{62}, 071503 (2000)},
\href{https://arxiv.org/abs/hep-ph/0005108}{\color{teal}{[arXiv:hep-ph/0005108 [hep-ph]]}}.
%902 citations counted in INSPIRE as of 28 Oct 2025

%\cite{Burkardt:2002hr}
\bibitem{Burkardt:2002hr}
M.~Burkardt,
Impact parameter space interpretation for generalized parton distributions,
\href{https://doi.org//10.1142/S0217751X03012370}{Int. J. Mod. Phys. A \textbf{18}, 173-208 (2003)},
\href{https://arxiv.org/abs/hep-ph/0207047}{\color{teal}{[arXiv:hep-ph/0207047 [hep-ph]]}}.
%844 citations counted in INSPIRE as of 28 Oct 2025

%\cite{Goeke:2001tz}
\bibitem{Goeke:2001tz}
K.~Goeke, M.~V.~Polyakov and M.~Vanderhaeghen,
Hard exclusive reactions and the structure of hadrons,
\href{https://doi.org//10.1016/S0146-6410(01)00158-2}{Prog. Part. Nucl. Phys. \textbf{47}, 401-515 (2001)},
\href{https://arxiv.org/abs/hep-ph/0106012}{\color{teal}{[arXiv:hep-ph/0106012 [hep-ph]]}}.
%1086 citations counted in INSPIRE as of 28 Oct 2025

%\cite{Diehl:2001pm}
\bibitem{Diehl:2001pm}
M.~Diehl,
Generalized parton distributions with helicity flip,
\href{https://doi.org//10.1007/s100520100635}{Eur. Phys. J. C \textbf{19}, 485-492 (2001)},
\href{https://arxiv.org/abs/hep-ph/0101335}{\color{teal}{[arXiv:hep-ph/0101335 [hep-ph]]}}.
%285 citations counted in INSPIRE as of 07 Dec 2025

%\cite{Diehl:2003ny}
\bibitem{Diehl:2003ny}
M.~Diehl,
Generalized parton distributions,
\href{https://doi.org//10.1016/j.physrep.2003.08.002}{Phys. Rept. \textbf{388}, 41-277 (2003)},
\href{https://arxiv.org/abs/hep-ph/0307382}{\color{teal}{[arXiv:hep-ph/0307382 [hep-ph]]}}.
%1627 citations counted in INSPIRE as of 28 Oct 2025

%\cite{Belitsky:2005qn}
\bibitem{Belitsky:2005qn}
A.~V.~Belitsky and A.~V.~Radyushkin,
Unraveling hadron structure with generalized parton distributions,
\href{https://doi.org//10.1016/j.physrep.2005.06.002}{Phys. Rept. \textbf{418}, 1-387 (2005)},
\href{https://arxiv.org/abs/hep-ph/0504030}{\color{teal}{[arXiv:hep-ph/0504030 [hep-ph]]}}.
%1164 citations counted in INSPIRE as of 28 Oct 2025


%\cite{ParticleDataGroup:2022pth}
\bibitem{ParticleDataGroup:2022pth}
R.~L.~Workman \textit{et al.} [Particle Data Group], Review of Particle Physics,
\href{https://doi.org//10.1093/ptep/ptac097}{PTEP \textbf{2022}, 083C01 (2022)}.
%5263 citations counted in INSPIRE as of 28 Oct 2025

%\cite{Ioffe:1981kw}
\bibitem{Ioffe:1981kw}
B.~L.~Ioffe,
Calculation of Baryon Masses in Quantum Chromodynamics,
\href{https://doi.org/10.1016/0550-3213(81)90259-5}{Nucl. Phys. B \textbf{188}, 317-341 (1981)}.
%doi:10.1016/0550-3213(81)90259-5
%1016 citations counted in INSPIRE as of 05 Apr 2025

%\cite{Aliev:2002ra}
\bibitem{Aliev:2002ra}
T.~M.~Aliev, A.~Ozpineci and M.~Savci,
Octet baryon magnetic moments in light cone QCD sum rules,
\href{https://doi.org/10.1103/PhysRevD.67.039901}{Phys. Rev. D \textbf{66}, 016002 (2002)},
%doi:10.1103/PhysRevD.67.039901
\href{https://arxiv.org/abs/hep-ph/0204035}{\color{teal}{[arXiv:hep-ph/0204035 [hep-ph]]}}.
%49 citations counted in INSPIRE as of 05 Apr 2025

%\cite{Azizi:2014yea}
\bibitem{Azizi:2014yea}
K.~Azizi and N.~Er,
Properties of nucleon in nuclear matter: once more,
\href{https://doi.org/10.1140/epjc/s10052-014-2904-5}{Eur. Phys. J. C \textbf{74}, 2904 (2014)},
%doi:10.1140/epjc/s10052-014-2904-5
\href{https://arxiv.org/abs/1401.1680}{\color{teal}{[arXiv:1401.1680 [hep-ph]]}}.
%20 citations counted in INSPIRE as of 05 Apr 2025

%\cite{Ozdem:2017jqh}
\bibitem{Ozdem:2017jqh}
U.~Ozdem and K.~Azizi,
Magnetic and quadrupole moments of the $Z_c(3900)$,
\href{https://doi.org//10.1103/PhysRevD.96.074030}{Phys. Rev. D \textbf{96}, 074030 (2017)},
\href{https://arxiv.org/abs/1707.09612}{\color{teal}{[arXiv:1707.09612 [hep-ph]]}}.
%28 citations counted in INSPIRE as of 06 Dec 2025 

%\cite{Azizi:2018duk}
\bibitem{Azizi:2018duk}
K.~Azizi, A.~R.~Olamaei and S.~Rostami,
Beautiful mathematics for beauty-full and other multi-heavy hadronic systems,
\href{https://doi.org//10.1140/epja/i2018-12595-1}{Eur. Phys. J. A \textbf{54}, 162 (2018)},
\href{https://arxiv.org/abs/1801.06789}{\color{teal}{[arXiv:1801.06789 [hep-ph]]}}.
%17 citations counted in INSPIRE as of 06 Dec 2025


%\cite{Azizi:2017ubq}
\bibitem{Azizi:2017ubq}
	K.~Azizi and N.~Er,
	X (3872): propagating in a dense medium,
	\href{https://doi.org//	10.1016/j.nuclphysb.2018.09.014}{	Nucl. Phys. B \textbf{936}, 151-168 (2018)},
	\href{https://arxiv.org/abs/1710.02806v3}{\color{teal}{	[arXiv:1710.02806 [hep-ph]]}}.
	%23 citations counted in INSPIRE as of 23 Oct 2025
%\cite{Faessler:2009xn}

%\cite{ParticleDataGroup:2024cfk}
\bibitem{ParticleDataGroup:2024cfk}
S.~Navas \textit{et al.} [Particle Data Group],
Review of particle physics,
\href{https://doi.org//10.1103/PhysRevD.110.030001}{Phys. Rev. D \textbf{110}, 030001 (2024)}.
%3121 citations counted in INSPIRE as of 11 Nov 2025


%\cite{Belyaev:1982sa}
\bibitem{Belyaev:1982sa}
V.~M.~Belyaev and B.~L.~Ioffe,
Determination of Baryon and Baryonic Resonance Masses from QCD Sum Rules. 1. Nonstrange Baryons,
\href{https://inis.iaea.org/records/r9jqb-2k079}{Sov. Phys. JETP \textbf{56}, 493-501 (1982)}.
%444 citations counted in INSPIRE as of 24 Oct 2025
%\cite{Belyaev:1982cd}
\bibitem{Belyaev:1982cd}
V.~M.~Belyaev and B.~L.~Ioffe,
Determination of the baryon mass and baryon resonances from the quantum-chromodynamics sum rule. Strange baryons,
\href{http://jetp.ras.ru/cgi-bin/dn/e_057_04_0716.pdf}{Sov. Phys. JETP \textbf{57}, 716-721 (1983)}.
%207 citations counted in INSPIRE as of 24 Oct 2025
%\cite{Aliev:2007pi}
\bibitem{Aliev:2007pi}
T.~M.~Aliev, K.~Azizi and A.~Ozpineci,
Light cone QCD sum rules analysis of the axial N -{\ensuremath{>}} Delta transition form-factors,
\href{https://doi.org//10.1016/j.nuclphysa.2007.11.006}{Nucl. Phys. A \textbf{799}, 105-126 (2008)},
\href{https://arxiv.org/abs/0707.1592}{\color{teal}{[arXiv:0707.1592 [hep-ph]]}}.
%19 citations counted in INSPIRE as of 24 Oct 2025
%\cite{DELPHI:1993ukk}
\bibitem{DELPHI:1993ukk}
P.~Abreu \textit{et al.} [DELPHI],
Determination of alpha-s from the scaling violation in the fragmentation functions in e+ e- annihilation,
\href{https://doi.org//10.1016/0370-2693(93)90587-8}{Phys. Lett. B \textbf{311}, 408-424 (1993)}.
%102 citations counted in INSPIRE as of 24 Oct 2025

%\cite{Pospelov:2005pr}
\bibitem{Pospelov:2005pr}
M.~Pospelov and A.~Ritz,
Electric dipole moments as probes of new physics,
\href{https://doi.org//10.1016/j.aop.2005.04.002}{Annals Phys. \textbf{318}, 119-169 (2005)},
\href{https://arxiv.org/abs/hep-ph/0504231}{\color{teal}{[arXiv:hep-ph/0504231 [hep-ph]]}}.
%1003 citations counted in INSPIRE as of 12 Dec 2025

%\cite{Wang:2018kto}
\bibitem{Wang:2018kto}
Q.~W.~Wang, S.~X.~Qin, C.~D.~Roberts and S.~M.~Schmidt,
Proton tensor charges from a Poincar{\'e}-covariant Faddeev equation,
\href{https://doi.org//10.1103/PhysRevD.98.054019}{Phys. Rev. D \textbf{98}, 054019 (2018)},
\href{https://arxiv.org/abs/1806.01287}{\color{teal}{[arXiv:1806.01287 [nucl-th]]}}.
%45 citations counted in INSPIRE as of 12 Dec 2025

%\cite{Cotogno:2019vjb}
\bibitem{Cotogno:2019vjb}
S.~Cotogno, C.~Lorc{\'e}, P.~Lowdon and M.~Morales,
Covariant multipole expansion of local currents for massive states of any spin,
\href{https://doi.org//10.1103/PhysRevD.101.056016}{Phys. Rev. D \textbf{101}, 056016 (2020)},
\href{https://arxiv.org/abs/1912.08749}{\color{teal}{[arXiv:1912.08749 [hep-ph]]}}.
%53 citations counted in INSPIRE as of 05 Dec 2025

\bibitem{Faessler:2009xn}
A.~Faessler, T.~Gutsche, M.~A.~Ivanov, J.~G.~Korner and V.~E.~Lyubovitskij,
Semileptonic decays of double heavy baryons in a relativistic constituent three-quark model,
\href{https://journals.aps.org/prd/abstract/10.1103/PhysRevD.80.034025}{Phys. Rev. D \textbf{80}, 034025 (2009)},
\href{https://arxiv.org/abs/0907.0563}{\color{teal}{[arXiv:0907.0563 [hep-ph]]}}.
%85 citations counted in INSPIRE as of 21 Oct 2025
%\cite{Agaev:2020zad}
\bibitem{Agaev:2020zad}
S.~Agaev, K.~Azizi and H.~Sundu,
Four-quark exotic mesons,
\href{https://doi.org//10.3906/fiz-2003-15}{Turk. J. Phys. \textbf{44}, 95 (2020)},
\href{https://arxiv.org/abs/2004.12079v1}{\color{teal}{[arXiv:2004.12079 [hep-ph]]}}.
%100 citations counted in INSPIRE as of 22 Oct 2025


\end{thebibliography}
\end{document}